\documentclass{IEEEtran}

\usepackage[T1]{fontenc}

\usepackage{standalone}
\usepackage{cite}
\usepackage{amsmath,amsfonts}
\usepackage{mathtools}
\usepackage{tabularray}
\usepackage{nicematrix}
\usepackage[shortmathrm]{physics-patch}
\usepackage{mathdots}
\usepackage[base]{medmath}
\usepackage{microtype}
\usepackage[dvipsnames,svgnames]{xcolor}
\usepackage{scontents}
\usepackage{tikz}
\usepackage{tikz-3dplot}
\usepackage{pgfplots}
\usepackage{flowchart}
\usepackage[hidelinks]{hyperref}
\usepackage[shortcuts]{extdash}
\usepackage{xoverline}
\ExplSyntaxOn
\NewDocumentCommand{\mytickprinter}{m}{
  \pgfmathprintnumberto[assume~math~mode=true]{#1}{\l_tmpa_tl}
  \tl_replace_once:Nnn \l_tmpa_tl {-} {\textminus}
  \tl_use:N \l_tmpa_tl
}
\ExplSyntaxOff

\pgfplotsset{
  2d-plot/.style={
    small,
    width=102pt,
    height=102pt,
    scale only axis,
    grid,
    grid style={black!15},
    cycle list name=color list,
    label style={inner sep=0.3ex},
    legend style={
      node font=\footnotesize,
      inner ysep=0.2ex,
      nodes={inner ysep=0.45ex},
    },
    legend cell align=left,
    xticklabel={\mytickprinter{\tick}},
    yticklabel={\mytickprinter{\tick}},
    tick label style={inner ysep=0.68ex},
    every axis plot/.append style={line join=round},
  },
  group2by1/.style={
    group style={
      group size=2 by 1,
      horizontal sep=1.4em,
      ylabels at=edge left,
      yticklabels at=edge left,
    },
    2d-plot,
    width=102pt,
    height=102pt,
    scale only axis,
    ylabel={Frequency [GHz]},
  },
  group1by2/.style={
    group style={
      group size=1 by 2,
      vertical sep=2.6ex,
      xlabels at=edge bottom,
      xticklabels at=edge bottom,
    },
    2d-plot,
    width=200pt,
    height=70pt,
    scale only axis,
    xlabel={Frequency [GHz]},
  },
}

\usetikzlibrary{external}
\usetikzlibrary{decorations.pathreplacing}
\usetikzlibrary{calligraphy}
\usetikzlibrary{patterns,patterns.meta}
\usetikzlibrary{graphs}
\usetikzlibrary{shapes.geometric,shapes.misc}
\usetikzlibrary{arrows.meta}
\usetikzlibrary{calc,backgrounds,fit}
\usetikzlibrary{pgfplots.groupplots}

\pgfplotsset{compat=1.18}

\NiceMatrixOptions{cell-space-limits=1pt}
\NiceMatrixOptions{xdots/horizontal-labels}

\UseTblrLibrary{booktabs}
\UseTblrLibrary{siunitx}

\xoverlineSetKeys{
  length* = 0.80,
  map = {
    A = { length*=0.68, shift*=2 },
    B = { shift*=1.5 },
    C = { shift*=1.2 },
    D = { shift*=1 },
    I = { shift*=1.45 },
    X = { shift*=1.2 },
    Y = { shift*= -0.35 },
    Z = { shift*=1.43 },
  },
}

\DeclareFontFamilySubstitution{TS1}{ptm}{qtm} 

\newsavebox{\boxA}

\newtoggle{pdf_instead_of_tikz}

\ExplSyntaxOn
\NewDocumentCommand{\hyphentominus}{m}{
  \tl_set:Nn \l_tmpa_tl {#1}
  \tl_replace_once:Nnn \l_tmpa_tl {-} {\textminus}
  \tl_use:N \l_tmpa_tl
}
\ExplSyntaxOff

\NewDocumentCommand{\todocite}{O{}}{\textcolor{purple}{\textit{cite #1}}}
\NewDocumentCommand{\draft}{+m}{\bgroup\color{gray!20!darkgray}#1\egroup}
\NewDocumentCommand{\explain}{+m}{\bgroup\itshape\color{gray}#1\egroup}
\NewDocumentCommand{\iws}{sm}{
  \unskip%
  \IfBooleanTF{#1}%
    {\hspace*{#2\fontdimen2\font plus #2\fontdimen3\font minus #2\fontdimen4\font}}%
    {\hspace{#2\fontdimen2\font plus #2\fontdimen3\font minus #2\fontdimen4\font}}%
  \ignorespaces%
}
\newcommand{\myeig}{{\smash[b]{\mathrm{eig}}}}

\newcommand{\figref}[1]{Fig.~\ref{#1}}

\newcommand{\?}{\iws*{0.83}}
\newcommand{\Real}{\opbraces{\operatorname{Re}}}
\newcommand{\Imag}{\opbraces{\operatorname{Im}}}
\newcommand{\sbar}[1]{\xoverline[1]{#1}}
\newcommand{\dbar}[1]{\xoverline[2]{#1}}

\NewDocumentEnvironment{revised}{}{\color{MediumBlue}}{}

\begin{document}
\title{Leaky-Wave Antenna Analysis using Multi-Modal Network Theory with Open Periodic Boundaries}

\author{
Oscar~Senlis,~\IEEEmembership{Student Member, IEEE}, 
John~N.~Le,~\IEEEmembership{Student Member, IEEE},
Anthony~Grbic,~\IEEEmembership{Fellow, IEEE},
Mauro~Ettorre,~\IEEEmembership{Fellow, IEEE},
Vincent~Laquerbe, 
David~Gonz\'alez-Ovejero,~\IEEEmembership{Senior Member, IEEE}
\thanks{
The work of O.~Senlis and D.~Gonz\'alez-Ovejero has been supported in part by the European Union through European Regional Development Fund (ERDF), Ministry of Higher Education and Research, CNRS, Brittany region, Conseils Départementaux d'Ille-et-Vilaine and Côtes-d'Armor, Rennes Métropole, and Lannion Trégor Communauté, through the CPER Project CyMoCod and in part by the Agence Nationale de la Recherche (ANR) through the AROMA project (reference ANR-22-CE24-0013).
The work of O.~Senlis was co-funded by CNES (Centre National d'\'Etudes Spatiales) and by AID (Defense Innovation Agency) from the French MINARM. The work of A.~Grbic was supported by US Air Force grant FA8650-22-D-5406 through a subcontract with Azimuth Corporation. The authors thank the University of Michigan for hosting O.~Senlis for a six-month academic exchange.}
\thanks{O. Senlis and D. González-Ovejero are with Univ Rennes CNRS, Institut d'\'Electronique et des Technologies du numéRique (IETR) -- UMR 6164, F-35000 Rennes, France.}
\thanks{A. Grbic and J. N. Le are with the Dept. of Electrical Engineering and Computer Science, University of Michigan, Ann Arbor, MI, USA.}
\thanks{M. Ettorre is with the Dept. of Electrical and Computer Engineering, Michigan State University, East Lansing, MI, USA.}
\thanks{V. Laquerbe is with Centre National d'\'Etudes Spatiales (CNES), Antenna Department, 31400 Toulouse, France.}
}

\maketitle

\begin{abstract}
This paper introduces two methods for analyzing periodic leaky-wave antennas (LWAs) within a new framework denoted as multi-modal network theory (MNT) with open periodic boundaries (OPBs). The approach is hybrid, combining analytical techniques with a commercial full-wave solver.
The first method computes the dispersion diagram of periodic LWAs. It is iterative and relies on the full-wave simulation of a single unit cell of a LWA, coupled with the analytical solution of an eigenvalue problem. This method effectively captures both the phase and attenuation constants of periodic LWAs while using fewer modes than previous methods with commercial frequency-domain solvers. The method is validated by computing the dispersion of classic LWA unit cells and comparing them to those obtained through full-wave simulations of the full-length antenna and other state-of-the-art methods.
The second, also based on OPB-MNT, focuses on LWA driven analysis.
Specifically, it determines the response of a unit cell to an incident plane wave. To validate this method, we compute the response of LWAs with different unit cell designs.
By comparing these results with the corresponding dispersion analysis, we highlight the differences between the driven case and eigenvalue problems and identify the scenarios in which the former is applicable.
\end{abstract}

\begin{IEEEkeywords}
  dispersion analysis, leaky-wave antenna, transfer matrix, multi-modal, simulation method.
\end{IEEEkeywords}

\section{Introduction}
\label{sec:introduction}

Dispersion diagrams plot the modal behavior of electromagnetic (EM) structures, showing how the complex wavenumbers of the supported modes vary with frequency. They provide essential insight into the modal behavior of EM band gap structures, leaky-wave antennas (LWAs), and metasurfaces. Consequently, numerous methods have been developed to compute dispersion diagrams. These methods include plane-wave expansion \cite{PhysRevLett.65.3152, PhysRevB.53.7134, PhysRevB.58.3721}, equivalent circuit modeling \cite{Palocz1970equivalent, Mesa2018unlocking}, spectral methods based on near-field measurements \cite{Sukhorukov:09}, and full-wave numerical techniques such as the Method of Moments (MoM) \cite{Mateo2012simple}.

In addition, multi-modal transfer matrix methods (MMTMM) have been used for dispersion analysis. This hybrid approach relies on full-wave simulations to compute the transmission\slash transfer matrix of the unit cell of a periodic structure, from which an eigenvalue equation is then solved \cite{Amari1998accurate, Apaydin2012Experimental}. It has been successfully used to compute the dispersion characteristics of one-dimensional \cite{Bandlow2008analysis, Bongard2009enhanced, Giusti2024eval}, two-dimensional \cite{Bagheriasl2019Bloch,Chen2021anisotropic} and three-dimensional closed or bounded periodic structures \cite{Giusti2022efficient, Giusti2022multimodal}.
The MMTMM has also been applied to unbounded or open-boundary structures \cite{garcia2025multimodal, Mesa2021simulation, Giusti2023linearized}, particularly for evaluating the attenuation constant of LWAs. However, the computed eigenvalues depend on the height of the wave port in the open region, necessitating time-consuming convergence studies to identify the optimal port height.

The overall contribution of this work is the development of a new framework for analyzing periodic open boundary structures based on multi-modal network theory \cite{Ranjbar:2017:Analysis, Mesa2021simulation, Young:2020:Metastructures, Alsolamy:2021:ModalNetwork}.
While belonging to the family of MMTMM \cite{garcia2025multimodal} and sharing similarities with this method, the OPB-MNT method yields accurate results while requiring fewer modes.
It is also largely independent of the simulation domain size of the open region. This new framework, referred to as MNT with open periodic boundaries (OPBs) is abbreviated as OPB\=/MNT. Based on this framework, the first concrete contribution is the introduction of a new method for computing the dispersion characteristics of periodic LWAs. The method is hybrid and iterative, consisting of two main steps that are repeated until the desired convergence threshold is reached. The first step involves using a commercial full-wave solver to simulate the unit cell with periodic boundary conditions (PBCs) applied to the open region. In the second step, post-processing is performed to solve a multi-modal eigenvalue problem for the guided field of the unit cell.

In addition to enabling the computation of the dispersion diagram of LWA, the OPB-MNT framework also allows for the analysis of the driven problem of a LWA illuminated by an incident plane wave \cite{Scarborough:2022:SimulatingSpace-Time}.
This incident plane wave problem is different from the eigen problem because the periodic phase delay is imposed across the unit cell by the incident plane wave.
The second concrete contribution of this work is a hybrid method based on OPB-MNT to analyze a periodic LWA under those circumstances and allows us to compare the unit cell's response in two different scenarios (driven and eigen problem) and examine the differences between them.


The paper is organized as follows. Section~\ref{sec:model} introduces the mathematical formulation of the OPB-MNT framework for analyzing the unit cells of periodic LWAs. Section~\ref{sec:dispersion} explains how the dispersion analysis is performed. It begins with a description of the required full-wave simulation setup, followed by a detailed presentation of the iterative algorithm that is employed. The dispersion for various LWAs are then computed using OPB-MNT and compared with those computed using other methods. Section~\ref{sec:receiving} is devoted to the driven analysis of LWAs. Numerical results for several examples are discussed and compared against solutions from the preceding dispersion\slash eigen problem. Finally, Section~\ref{sec:conclusion} concludes the paper by highlighting the advantages of the proposed method.

\section{Circuit Modeling of Periodic Leaky-Wave Antennas}
\label{sec:model}

\begin{figure}
  \centering%
  \iftoggle{pdf_instead_of_tikz}{%
    \includegraphics{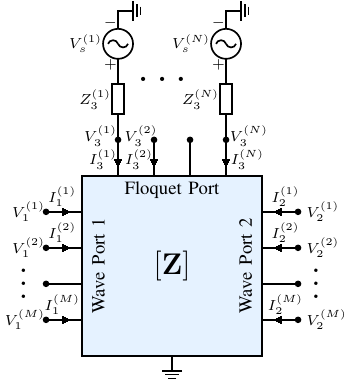}%
  }{%
    \tikzsetnextfilename{circuit}%
    \begin{tikzpicture}[scale=2.54]%
\ifx\dpiclw\undefined\newdimen\dpiclw\fi
\global\def\dpicdraw{\draw[line width=\dpiclw]}
\global\def\dpicstop{;}
\dpiclw=0.8bp
\dpiclw=0.8bp
\tikzset{every node/.style={node font=\small}}
\dpicdraw[fill={rgb,1:red,0.90000;green,0.95000;blue,1.00000}](0.256,-0.653) rectangle (1.456,0.547)\dpicstop
\draw (0.856,-0.053) node{{\Large$[$\raisebox{-0.1ex}{\Large$\mathbf{Z}$}$]$}};
\dpicdraw (0.256,0.307)
 --(0.016,0.307)\dpicstop
\dpicdraw[fill=black](0.016,0.307) circle (0.006299in)\dpicstop
\dpicdraw (0.256,0.067)
 --(0.016,0.067)\dpicstop
\dpicdraw[fill=black](0.016,0.067) circle (0.006299in)\dpicstop
\dpicdraw (0.256,-0.173)
 --(0.016,-0.173)\dpicstop
\dpicdraw[fill=black](0.016,-0.173) circle (0.006299in)\dpicstop
\dpicdraw (0.256,-0.413)
 --(0.016,-0.413)\dpicstop
\dpicdraw[fill=black](0.016,-0.413) circle (0.006299in)\dpicstop
\dpicdraw (0.496,0.547)
 --(0.496,0.787)\dpicstop
\dpicdraw[fill=black](0.496,0.787) circle (0.006299in)\dpicstop
\dpicdraw (0.736,0.547)
 --(0.736,0.787)\dpicstop
\dpicdraw[fill=black](0.736,0.787) circle (0.006299in)\dpicstop
\dpicdraw (0.976,0.547)
 --(0.976,0.787)\dpicstop
\dpicdraw[fill=black](0.976,0.787) circle (0.006299in)\dpicstop
\dpicdraw (1.216,0.547)
 --(1.216,0.787)\dpicstop
\dpicdraw[fill=black](1.216,0.787) circle (0.006299in)\dpicstop
\dpicdraw (1.456,0.307)
 --(1.696,0.307)\dpicstop
\dpicdraw[fill=black](1.696,0.307) circle (0.006299in)\dpicstop
\dpicdraw (1.456,0.067)
 --(1.696,0.067)\dpicstop
\dpicdraw[fill=black](1.696,0.067) circle (0.006299in)\dpicstop
\dpicdraw (1.456,-0.173)
 --(1.696,-0.173)\dpicstop
\dpicdraw[fill=black](1.696,-0.173) circle (0.006299in)\dpicstop
\dpicdraw (1.456,-0.413)
 --(1.696,-0.413)\dpicstop
\dpicdraw[fill=black](1.696,-0.413) circle (0.006299in)\dpicstop
\draw (0.306,-0.053) node[right=-2bp]{\rotatebox{90}{Wave Port 1}};
\draw (1.406,-0.053) node[left=-2bp]{\rotatebox{90}{Wave Port 2}};
\draw (0.856,0.522) node[below=-2bp]{{Floquet Port}};
\dpicdraw (0.856,-0.653)
 --(0.856,-0.753)\dpicstop
\dpicdraw (0.922667,-0.753)
 --(0.789333,-0.753)\dpicstop
\dpicdraw (0.900444,-0.778)
 --(0.811556,-0.778)\dpicstop
\dpicdraw (0.884571,-0.803)
 --(0.827429,-0.803)\dpicstop
\tikzset{every node/.style={node font=\scriptsize}}
\draw (0.016,0.307) node[left=-2bp]{{$V^{(1)}_1$}};
\draw (0.016,0.067) node(V12)[left=-2bp]{{$V^{(2)}_1$}};
\draw (0.016,-0.413) node(V1n)[left=-2bp]{{$V^{(M)}_1$}};
\filldraw[line width=0bp](0.1244,0.282)
 --(0.1744,0.307)
 --(0.1244,0.332) --cycle\dpicstop
\dpicdraw (0.016,0.307)
 --(0.161977,0.307)\dpicstop
\draw (0.1244,0.332) node[above=-2bp]{$I^{(1)}_1$};
\filldraw[line width=0bp](0.1244,0.042)
 --(0.1744,0.067)
 --(0.1244,0.092) --cycle\dpicstop
\dpicdraw (0.016,0.067)
 --(0.161977,0.067)\dpicstop
\draw (0.1244,0.092) node[above=-2bp]{$I^{(2)}_1$};
\filldraw[line width=0bp](0.1244,-0.438)
 --(0.1744,-0.413)
 --(0.1244,-0.388) --cycle\dpicstop
\dpicdraw (0.016,-0.413)
 --(0.161977,-0.413)\dpicstop
\draw (0.1244,-0.388) node[above=-2bp]{$I^{(M)}_1$};
\draw (-0.134,-0.1058) node{\LARGE $\vdots$};
\draw (1.746,0.307) node[right=-2bp]{{$V^{(1)}_2$}};
\draw (1.746,0.067) node(V22)[right=-2bp]{{$V^{(2)}_2$}};
\draw (1.746,-0.413) node(V2n)[right=-2bp]{{$V^{(M)}_2$}};
\filldraw[line width=0bp](1.5876,0.332)
 --(1.5376,0.307)
 --(1.5876,0.282) --cycle\dpicstop
\dpicdraw (1.696,0.307)
 --(1.550023,0.307)\dpicstop
\draw (1.6126,0.332) node[above=-2bp]{$I^{(1)}_2$};
\filldraw[line width=0bp](1.5876,0.092)
 --(1.5376,0.067)
 --(1.5876,0.042) --cycle\dpicstop
\dpicdraw (1.696,0.067)
 --(1.550023,0.067)\dpicstop
\draw (1.6126,0.092) node[above=-2bp]{$I^{(2)}_2$};
\filldraw[line width=0bp](1.5876,-0.388)
 --(1.5376,-0.413)
 --(1.5876,-0.438) --cycle\dpicstop
\dpicdraw (1.696,-0.413)
 --(1.550023,-0.413)\dpicstop
\draw (1.6126,-0.388) node[above=-2bp]{$I^{(M)}_2$};
\draw (1.816,-0.1058) node{\LARGE $\vdots$};
\draw (0.496,0.812) node[left=-2bp]{{$V^{(1)}_3$}};
\draw (0.766,0.812) node[left=-2bp]{{$V^{(2)}_3$}};
\draw (1.216,0.812) node[right=-2bp]{{$\,V^{(N)}_3$}};
\filldraw[line width=0bp](0.471,0.657)
 --(0.496,0.607)
 --(0.521,0.657) --cycle\dpicstop
\dpicdraw (0.496,0.787)
 --(0.496,0.619423)\dpicstop
\draw (0.496,0.657) node[left=-2bp]{$I^{(1)}_3$};
\filldraw[line width=0bp](0.711,0.657)
 --(0.736,0.607)
 --(0.761,0.657) --cycle\dpicstop
\dpicdraw (0.736,0.787)
 --(0.736,0.619423)\dpicstop
\draw (0.736,0.657) node[left=-2bp]{$I^{(2)}_3$};
\filldraw[line width=0bp](1.191,0.657)
 --(1.216,0.607)
 --(1.241,0.657) --cycle\dpicstop
\dpicdraw (1.216,0.787)
 --(1.216,0.619423)\dpicstop
\draw (1.241,0.657) node[right=-2bp]{$I^{(N)}_3$};
\dpicdraw (0.496,0.787)
 --(0.496,0.957)\dpicstop
\dpicdraw (0.496,1.157)
 --(0.456,1.157)
 --(0.456,0.957)
 --(0.536,0.957)
 --(0.536,1.157)
 --(0.496,1.157)\dpicstop
\dpicdraw (0.496,1.157)
 --(0.496,1.327)\dpicstop
\draw (0.456,1.057) node[left=-2bp]{$ Z^{(1)}_3$};
\dpicdraw (0.496,1.327)
 --(0.496,1.327)\dpicstop
\dpicdraw (0.496,1.427) circle (0.03937in)\dpicstop
\dpicdraw (0.496,1.427)
 ..controls (0.491217,1.441348) and (0.47779,1.451025)
 ..(0.462667,1.451025)
 ..controls (0.447543,1.451025) and (0.434116,1.441348)
 ..(0.429333,1.427)\dpicstop
\dpicdraw (0.496,1.427)
 ..controls (0.500783,1.412652) and (0.51421,1.402975)
 ..(0.529333,1.402975)
 ..controls (0.544457,1.402975) and (0.557884,1.412652)
 ..(0.562667,1.427)\dpicstop
\dpicdraw (0.496,1.527)
 --(0.496,1.527)\dpicstop
\draw (0.496,1.327) node[below left=-2bp]{$ +$};
\draw (0.396,1.427) node[left=-2bp]{$ V^{(1)}_s$};
\draw (0.496,1.527) node[above left=-2bp]{$ -$};
\dpicdraw (0.496,1.527)
 --(0.496,1.647)\dpicstop
\dpicdraw (0.496,1.641444)
 --(0.496,1.652556)\dpicstop
\dpicdraw (0.496,1.647)
 --(0.596,1.647)\dpicstop
\dpicdraw (0.596,1.713667)
 --(0.596,1.580333)\dpicstop
\dpicdraw (0.621,1.691444)
 --(0.621,1.602556)\dpicstop
\dpicdraw (0.646,1.675571)
 --(0.646,1.618429)\dpicstop
\dpicdraw (1.216,0.787)
 --(1.216,0.957)\dpicstop
\dpicdraw (1.216,1.157)
 --(1.176,1.157)
 --(1.176,0.957)
 --(1.256,0.957)
 --(1.256,1.157)
 --(1.216,1.157)\dpicstop
\dpicdraw (1.216,1.157)
 --(1.216,1.327)\dpicstop
\draw (1.176,1.057) node[left=-2bp]{$ Z^{(N)}_3$};
\dpicdraw (1.216,1.327)
 --(1.216,1.327)\dpicstop
\dpicdraw (1.216,1.427) circle (0.03937in)\dpicstop
\dpicdraw (1.216,1.427)
 ..controls (1.211217,1.441348) and (1.19779,1.451025)
 ..(1.182667,1.451025)
 ..controls (1.167543,1.451025) and (1.154116,1.441348)
 ..(1.149333,1.427)\dpicstop
\dpicdraw (1.216,1.427)
 ..controls (1.220783,1.412652) and (1.23421,1.402975)
 ..(1.249333,1.402975)
 ..controls (1.264457,1.402975) and (1.277884,1.412652)
 ..(1.282667,1.427)\dpicstop
\dpicdraw (1.216,1.527)
 --(1.216,1.527)\dpicstop
\draw (1.216,1.327) node[below left=-2bp]{$ +$};
\draw (1.116,1.427) node[left=-2bp]{$ V^{(N)}_s$};
\draw (1.216,1.527) node[above left=-2bp]{$ -$};
\dpicdraw (1.216,1.527)
 --(1.216,1.647)\dpicstop
\dpicdraw (1.216,1.641444)
 --(1.216,1.652556)\dpicstop
\dpicdraw (1.216,1.647)
 --(1.316,1.647)\dpicstop
\dpicdraw (1.316,1.713667)
 --(1.316,1.580333)\dpicstop
\dpicdraw (1.341,1.691444)
 --(1.341,1.602556)\dpicstop
\dpicdraw (1.366,1.675571)
 --(1.366,1.618429)\dpicstop
\draw (0.814,1.192) node{\huge $\dots$};
\end{tikzpicture}%
%
  }%
  \caption{A 3-port multi-modal impedance network representation of a LWA unit cell. The wave ports characterize the guided EM field, while the Floquet port characterizes the incident and radiating EM field.}
  \label{fig:circuit}
\end{figure}

The unit cell of a LWA can be modeled as a multi-modal 3\=/port network, as shown in \figref{fig:circuit}. The first two multi-modal ports, each supporting $M$ modes, model the guided EM fields of the periodic structure. The third multi-modal port, supporting $N$ modes, represents a Floquet port that accounts for the scattered\slash radiated and incident EM fields. The multi-modal total voltage $\sbar{V}_i$ and current $\sbar{I}_i$ vectors are related through the following $(2M\mathinner+N)\times(2M\mathinner+N)$ Z\=/parameter block matrix:
\vspace{-0.3ex}
\begin{equation}    \label{eq:z-matrix}
  \begin{bNiceMatrix}
    \sbar{V}_1 \vphantom{\dbar{Z}_{12}} \\
    \sbar{V}_2 \vphantom{\dbar{Z}_{12}} \\
    \sbar{V}_3 \vphantom{\dbar{Z}_{12}}
  \end{bNiceMatrix}
  =
\tikzset{nicematrix/brace/.style = {decoration = {amplitude=0.4em, calligraphic brace, raise=-0.65ex}, line width = 0.07em, decorate}}
\setlength{\arraycolsep}{4pt}
  \begin{bNiceMatrix}[margin, first-row]
    \Hbrace{2}{2M} & \Hbrace{1}{N} \\
    \Block[borders={right,bottom,tikz=dotted}]{2-2}{}\dbar{Z}_{11} & \dbar{Z}_{12} & \dbar{Z}_{13} \\
    \dbar{Z}_{21} & \dbar{Z}_{22} & \dbar{Z}_{23} \\
    \dbar{Z}_{31} & \dbar{Z}_{32} & \Block[borders={left,top,tikz=dotted}]{1-1}{}\dbar{Z}_{33}
  \end{bNiceMatrix}
\tikzset{nicematrix/brace/.style = {decoration = {amplitude=0.34em, calligraphic brace, raise=-0.18em}, line width = 0.07em, decorate}}
  \begin{bNiceMatrix}[last-col]
    \sbar{I}_1 \vphantom{\dbar{Z}_{12}} & \Vbrace[shorten-start=2pt]{2}{2M} \\
    \sbar{I}_2 \vphantom{\dbar{Z}_{12}} & \\
    \sbar{I}_3 \vphantom{\dbar{Z}_{12}} & \Vbrace[shorten-start=0.5pt]{1}{N}
  \end{bNiceMatrix}
\end{equation}

Assuming one-dimensional spatial periodicity along the $x$\=/axis (as defined in \figref{fig:3d_view}), the guided total voltages and currents at wave ports 1 and 2 can be related by the Floquet theorem through the complex propagation constant $\kappa=\beta-j\alpha$ and the spatial period $d$, yielding:
\begin{subequations}\label{eq:VIejkd}
\begin{align}
  \sbar{V}_1 &=  \sbar{V}_2 e^{j\kappa d} \\
  \sbar{I}_1 &= -\sbar{I}_2 e^{j\kappa d}
\end{align}
\end{subequations}
Given the excitation vector $\sbar{V}_{\rms}$ and characteristic impedances of the Floquet modes given by $\dbar{Z}_3$, the total voltages and currents at the Floquet port are related through:
\begin{equation} \label{eq:VI-floquet}
   \sbar{V}_3 = \sbar{V}_{\rms} -\dbar{Z}_3 \sbar{I}_3
\end{equation}
where,
\begin{align}
  \sbar{V}_{\rms} &=                     \PTqty\big[ V^{(1)}_{\rms}\!, V^{(2)}_{\rms}\!, \dots, V^{(N)}_{\rms}]^{\rmT} \\
  \dbar{Z}_3      &= \operatorname{diag} \PTqty\big[ Z^{(1)}_3\!, Z^{(2)}_3 \!,\dots, Z^{(N)}_3 ]
\end{align}

Finally, by combining \eqref{eq:z-matrix}, \eqref{eq:VIejkd}, and \eqref{eq:VI-floquet}, we obtain the following matrix equation, where $\vb{I}_{M}$ denotes the identity matrix of size $M$ and $\vb{0}_{N\!,M}$ is the null matrix of size $N\times\mspace{0.8mu}M$.
\begin{equation}    \label{eq:matrix-driven}
\setlength{\arraycolsep}{6pt}
  \begin{bmatrix*}[r]
    -e^{j\kappa d}\vb{I}_M & \dbar{Z}_{12}\mathinner-e^{j\kappa d}\dbar{Z}_{11} &                      \dbar{Z}_{13} \\
                 -\vb{I}_M & \dbar{Z}_{22}\mathinner-e^{j\kappa d}\dbar{Z}_{21} &                      \dbar{Z}_{23} \\
            \vb{0}_{N\!,M} & \dbar{Z}_{32}\mathinner-e^{j\kappa d}\dbar{Z}_{31} & \dbar{Z}_3\mathinner+\dbar{Z}_{33} \\
  \end{bmatrix*}\mspace{-3mu}
  \begin{bmatrix}
    \sbar{V}_2 \vphantom{\dbar{Z}_{12}} \\
    \sbar{I}_2 \vphantom{\dbar{Z}_{12}} \\
    \sbar{I}_3 \vphantom{\dbar{Z}_{12}} \\
  \end{bmatrix}
  {=}
  \begin{bmatrix}
    0               \vphantom{\dbar{Z}_{12}} \\
    0               \vphantom{\dbar{Z}_{12}} \\
    \sbar{V}_{\rms} \vphantom{\dbar{Z}_{12}} \\
  \end{bmatrix}
\end{equation}
This matrix equation is the basis of the OPB-MNT analysis presented in this paper. It is used to analyze periodic LWAs both in the absence of an incident wave (eigenvalue problem) or in its presence (driven problem).

\section{Dispersion Analysis using OPB-MNT}    \label{sec:dispersion}

By suppressing any incident wave in the problem and setting the source vector $\sbar{V}_{\rms}=\text{0}$, \eqref{eq:matrix-driven} is transformed into an eigenvalue problem. Non-trivial solutions are found by solving for the complex wavenumber $\kappa$ in the following equation:
\begin{equation}    \label{eq:matrix-eigen}
\hspace{\widthof{\eqref{eq:matrix-eigen}}*\real{-0.25}}
\setlength{\arraycolsep}{6pt}
  \det(
  \begin{matrix*}[r]
    -e^{j\kappa d}\vb{I}_M & \dbar{Z}_{12}\mathinner-e^{j\kappa d}\dbar{Z}_{11} &                      \dbar{Z}_{13} \\
                 -\vb{I}_M & \dbar{Z}_{22}\mathinner-e^{j\kappa d}\dbar{Z}_{21} &                      \dbar{Z}_{23} \\
            \vb{0}_{N\!,M} & \dbar{Z}_{32}\mathinner-e^{j\kappa d}\dbar{Z}_{31} & \dbar{Z}_3\mathinner+\dbar{Z}_{33} \\
  \end{matrix*}
  ) = 0
\end{equation}


\begin{figure}
  \centering%
  \iftoggle{pdf_instead_of_tikz}{%
    \includegraphics{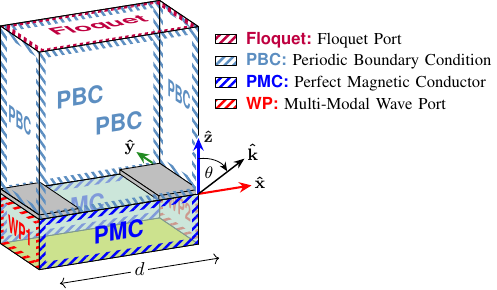}%
  }{%
    \tikzsetnextfilename{3d_main}%
    \input{figure/3d_view/3d_main.tex}%
  }%
  \caption{%
    A perspective view of the unit cell of a LWA. Boundary conditions and ports are shown on the exterior faces.
    The phase delay between the two PBCs transverse to $\vu{x}$ is equal to $\kappa d$, while the phase delay is set to 0 for the PBCs transverse to $\vu{y}$.
  }
  \label{fig:3d_view}
\end{figure}

In order to solve \eqref{eq:matrix-eigen}, the multi-modal impedance matrix representing the periodic LWA unit cell is retrieved using a commercial full-wave EM solver. We selected COMSOL \cite{COMSOL} because it allows a complex phase delay to be imposed across the PBCs, which is necessary for obtaining accurate dispersion results within the proposed framework. \figref{fig:3d_view} depicts an example LWA unit cell, consisting of metallic strips on a grounded dielectric slab, together with the simulation setup used. 
The boundaries delimiting the direction of propagation (along the $x$-axis), are divided into two parts.
The first part is dedicated to the wave ports, WP1 and WP2, whose location and extent are chosen to ensure that they excite propagating modes within the structure.
This configuration allows the fields at each wave port to be expanded in terms of a finite number of port modes.
\figref{fig:3d_view} illustrates an example of a periodic metal strip grating LWA, in which the wave ports are defined to excite parallel plate waveguide (PPW) modes.
Accordingly, they are located between the bottom and top metallizations. 
The remaining parts of these boundaries are assigned PBCs, which impose a \emph{complex} phase delay, corresponding to the wavenumber $\kappa$, across the open structure.
As it will be shown later, the wave ports and the PBCs can also be in direct contact.
The top of the domain is terminated by a Floquet port. In this particular example, to account for the $y$ invariance in this 2D problem, perfect magnetic conductor (PMC) boundaries are applied at the two $y$ boundaries of the lower region of the unit cell, while an additional PBC pair with zero phase delay is placed in the upper region. Therefore, the width of the unit cell must be made small enough such that no modes with $y$ dependence are excited.

Since the complex wavenumber $\kappa$ used to define the PBC is unknown at first, the simulation is performed using $\kappa_0$, an initial guess for $\kappa$. Because $\kappa_0$ is arbitrarily chosen, the initial phase delay imposed on the EM field across the PBCs is incorrect. Therefore, solving \eqref{eq:matrix-eigen} will yield an incorrect complex wavenumber, denoted $\kappa_0^\myeig$. The goal of the OPB-MNT method is to iteratively minimize the absolute difference $\abs{\Delta\kappa}$ between these two complex wavenumbers: the one imposed by the PBC and the one obtained by solving the eigenvalue problem in \eqref{eq:matrix-eigen} for the EM field at the wave ports. The complex wavenumber imposed across the PBCs is considered correct when it matches the eigenvalue obtained from \eqref{eq:matrix-eigen}. The iterative process terminates once $\abs{\Delta\kappa}$ falls below a prescribed convergence threshold $\epsilon_\kappa$.

Alternatively, instead of solving the determinantal equation \eqref{eq:matrix-eigen}, the wavenumber can be computed by solving the eigenvalue problem derived from the multi-modal transfer matrix (ABCD) of the unit cell \cite{Pozar:2011:MicroEngin}:
\begin{equation}    \label{eq:matrix-abcd}
  \begin{bmatrix}
    \dbar{A} & \dbar{B} \\
    \dbar{C} & \dbar{D}
  \end{bmatrix}
  \begin{bmatrix}
    \sbar{V}_1 \vphantom{\dbar{A}} \\ \sbar{I}_1 \vphantom{\dbar{C}}
  \end{bmatrix}
   = e^{j\kappa d}
  \begin{bmatrix}
    \sbar{V}_1 \vphantom{\dbar{A}} \\ \sbar{I}_1 \vphantom{\dbar{C}}
  \end{bmatrix}
\end{equation}
where $\dbar{A}$, $\dbar{B}$, $\dbar{C}$, $\dbar{D}$ are $M \times M$ block transfer matrix parameters.
The multi-modal ABCD parameters can be computed from the S\=/parameters following the derivation detailed in \cite{Mesa2021simulation} or \cite{Coves2012fullwave}.
In this case, we terminate the Floquet port with the characteristic impedance of the Floquet modes to recover the correct S\=/parameters. Alternatively, we can replace the Floquet port in the full-wave simulation setup (\figref{fig:3d_view}) by a radiation boundary to directly recover the same $2M\times\mspace{2mu}2M$ ABCD matrix.
Finally, it should be noted that solutions of \eqref{eq:matrix-eigen} and the $2M$ eigenvalues of the multi-modal unit cell's transfer matrix are identical.

\subsection{Description of the algorithm}

\begin{figure}
  \centering%
  \iftoggle{pdf_instead_of_tikz}{%
    \includegraphics{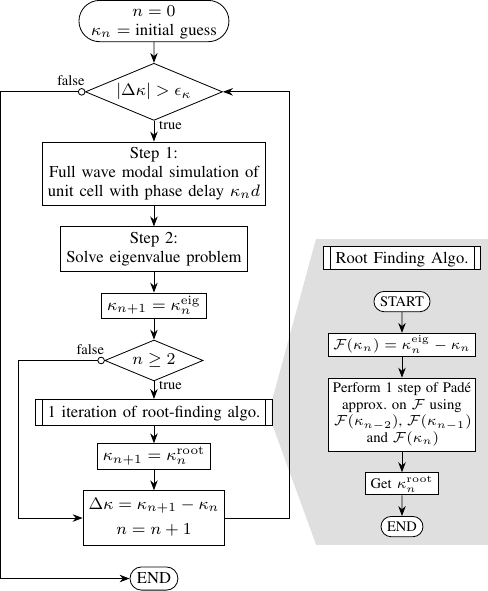}%
  }{%
    \tikzsetnextfilename{flow-chart3}%
    \begin{tikzpicture}[
  >=Stealth,
  graphs/every graph/.style={
    grow down sep,
    nodes={
      node font=\footnotesize,
      minimum height=3.10ex,
      draw,
    }
  },
  left lbl/.style={
    label position=left,
    node font=\scriptsize, inner xsep=0, inner ysep=0.9ex,anchor=south east
  },
  down lbl/.style={
    label position=below,
    node font=\scriptsize, inner ysep=0,anchor=north west
  },
  right loop/.style={to path={-- ++(#1,0) |- (\tikztotarget)}},
  left loop/.style={to path={-- pic[pos=0] {code={\draw[fill=white] (-1.6pt,0) circle[radius=1.6pt];}} ++(-#1,0) |- (\tikztotarget)}},
  io node/.style={trapezium, trapezium left angle=70, trapezium right angle=110, inner xsep=0.2em},
  smashb/.style={execute at begin node={\setbox0=\hbox\bgroup},execute at end node={\egroup\smash[b]{\box0}\vrule width0pt height0sp depth0.11ex}},
  decision/.style={diamond,aspect=2.4,inner ysep=0.3ex,label={[down lbl]true},label={[left lbl]false}},
]

\pgfmathsetlengthmacro\loopshift{2.6cm}
\pgfmathsetlengthmacro\loopshiftB{2.3cm}
\renewcommand{\band}{3pt} 

\graph {
  start / "$n=0$\\$\kappa_n={}$initial guess" [rounded rectangle,align=center,smashb]
  -> while / $\abs{\Delta\kappa} > \epsilon_\kappa$ [decision]
  -> "Step 1:\\Full wave modal simulation of\\unit cell with phase delay $\kappa_n d$" [align=center]
  -> "Step 2:\\Solve eigenvalue problem" [align=center]
  -> "$\kappa_{n+1}=\kappa_n^{\smash{\mathrm{eig}}\xmathstrut{-0.45}}$" [smashb,minimum height=3.4ex]
  -> nGeq2 / "$n\geq2$" [decision]
  -> rfg/"1 iteration of root-finding algo." [predproc,smashb,outer sep=-1.2pt,minimum height=3.4ex]
  -> "$\kappa_{n+1}=\kappa_n^{\smash{\mathrm{root}}\xmathstrut{-0.45}}$" [smashb,minimum height=3.4ex]
  -> last/
    "$
      \begin{gathered}
        \Delta\kappa = \kappa_{n+1} - \kappa_n\\[-0.0\jot]
        n=n+1
      \end{gathered}
    $"
  ->[right loop=\loopshiftB] while
  -!- END[rounded rectangle];

  while ->[left loop=\loopshift] END;
  nGeq2 ->[left loop=\loopshiftB] last;
};

\path[xshift=4.2cm,yshift=-4.17cm] graph[nodes={node font=\scriptsize,fill=white}]{
  TITLE/Root Finding Algo.[node font=\footnotesize, predproc,smashb]

  -!- START [rounded rectangle]
  -> "$\mathcal{F}(\kappa_n) = \kappa_n^{\smash{\mathrm{eig}}} - \kappa_n$" [text depth=0.4ex]
  -> "Perform 1 step of Pad\'e approx.\ on $\mathcal{F}$ using $\mathcal{F}(\kappa_{n-2})$, $\mathcal{F}(\kappa_{n-1})$ and $\mathcal{F}(\kappa_n)$" [text width=9.5em,align=center]
  -> "Get $\kappa_n^{\smash{\mathrm{root}}}$" [smashb,minimum height=3.4ex]
  -> END [rounded rectangle]
};

\begin{scope}[on background layer]
  \node[fill=lightgray!50, draw=lightgray!50, fit=(TITLE) (END)] (bg) {};
  \fill[lightgray!50] (rfg.north east) ++(0.5\band+\pgflinewidth,0) -- (bg.north west) -- (bg.south west) -- ($(rfg.south east)+(0.5\band+\pgflinewidth,0)$) -- cycle;
\end{scope}

\end{tikzpicture}%
  }%
  \caption{Flow chart of the iterative method used to compute the complex
  wave number at a particular frequency point using the OPB-MNT.}
  \label{fig:flow-chart}
\end{figure}

This section describes the iterative method for finding $\kappa$, as illustrated in \figref{fig:flow-chart}. It consists of two steps (full-wave simulation and the solution of an eigenvalue problem) that are alternated to obtain an accurate estimate of $\kappa$.

The first step is to perform a full-wave simulation of the LWA unit cell under study using a commercial full-wave solver to recover the network description of the unit cell. A PBC is applied across the unit cell (see \figref{fig:3d_view}), with $\kappa_0$ as an initial guess for the wavenumber $\kappa$ at iteration $n=\text{0}$.

In the second step, post-processing is performed. The multi-modal Z\=/parameters (referenced to the wave ports that terminate the unit cell) are obtained from full-wave simulation and used to set up the eigenvalue problem. This can be done using either the impedance parameters, as given in \eqref{eq:matrix-eigen}, or the transfer matrix, as given in \eqref{eq:matrix-abcd}.
To solve \eqref{eq:matrix-eigen}, we utilize a complex root-finding procedure, specifically the type~II Padé approximation-based algorithm in \cite{galdi_simple_2000}. The initial estimate \(\kappa_0\) can be used as a starting point in the root-finding process to determine the solution \(\kappa_0^\myeig\). Alternatively, if we employ the transfer matrix method, solving the eigenvalue problem in \eqref{eq:matrix-abcd} will produce \(2M\) eigenvalues $e^{j\kappa d}$, each corresponding to a complex wavenumber $\kappa$.
Eigenvalues corresponding to eigenmodes with power flow in the forward direction will have magnitudes above one, or, equivalently, a positive attenuation constant ($\alpha$). It should be noted that, to ensure convergence using the iterative method, the eigenvalue corresponding to the same eigenvector of interest must be chosen in each iteration. If multiple eigenvalues\slash eigenvectors are of interest, the iterative method must be performed separately for each, as illustrated in Section~\ref{Sec:dual-mode}.

At this point, the next iteration ($n=\text{1}$) of the algorithm starts with an updated estimate of the complex wavenumber $\kappa_1=\kappa_0^\myeig$. This iterative process then continues until a specified convergence threshold, $\epsilon_\kappa$, is satisfied.
Although this simple fixed-point iteration typically converges, its robustness can be improved by introducing an additional step before updating the value of $\kappa_n$ to be used in the next full-wave simulation.

This additional step consists of applying a root-finding procedure to the function $\mathcal{F}$ defined by $\kappa_n \mapsto \kappa_n^\myeig-\kappa_n$, which represents the difference between the complex wavenumber imposed in the current simulation iteration ($\kappa_n$) and the wavenumber retrieved from the eigenvalue problem ($\kappa_n^\myeig$). The root of this function, $\kappa^{\mathrm{root}}$, therefore corresponds to the correct wavenumber $\kappa$ as it implies $\kappa^{\mathrm{root}}=\kappa^\myeig$.
The same type~II Padé approximation\-/based root-finding algorithm \cite{galdi_simple_2000} is reused for this step. Since this algorithm requires three previous values of $\kappa_n$, the decision box initially bypasses the root-finding step when $n < \text{2}$, as illustrated in the flowchart of \figref{fig:flow-chart}. However, once $n \geq \text{2}$, the algorithm proceeds with the root-finding process described above, as illustrated in the highlighted gray area in \figref{fig:flow-chart}.
In summary, as the iterative process progresses, the value of $\kappa_n^{\mathrm{root}}$, obtained from the root-finding algorithm is used in the subsequent iteration instead of $\kappa_n^\myeig$ to define the phase delay between PBCs in the next full-wave simulation.

\subsection{Numerical Results}

\begin{figure}
  \centering%
  \iftoggle{pdf_instead_of_tikz}{%
    \includegraphics{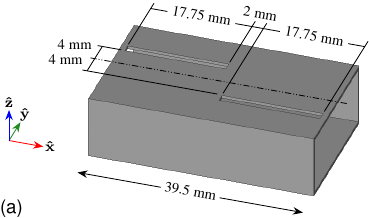}%
  }{%
    \tikzsetnextfilename{waveguide}%
    \input{figure/waveguide/waveguide.tex}%
  }%
  \vskip10pt%
  \iftoggle{pdf_instead_of_tikz}{%
    \includegraphics{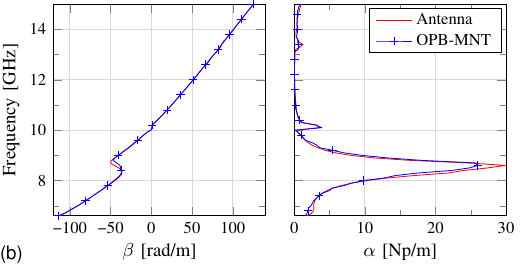}%
  }{%
    \tikzsetnextfilename{disp_waveguide}%
    \begin{tikzpicture}

\pgfmathsetmacro{\betaPi}{pi/39.5e-3}

\begin{groupplot}[
  group2by1,
  ymin=6.6,
  ymax=15,
  minor y tick num=1,
]

\nextgroupplot[ 
  xlabel={$\beta$ [rad/m]},
  xmin = -120,
  xmax = 140,
]

\addplot+[
  no markers,
  line join=round,
  x filter/.expression={x < -120 ? x+4*\betaPi : x},
] table [
  x index=3, 
  y index=1,
  y expr=\thisrowno{1}*1e-9,
] {\CurrentFilePath/ref.dat};

\addplot+[
  mark=+,
  mark repeat=6,
] table [
  x index=3, 
  y index=1,
  y expr=\thisrowno{1}*1e-9,
] {\CurrentFilePath/optmm.dat};

\nextgroupplot[ 
  xlabel={$\alpha$ [Np/m]},
  xmin=0,
  xmax=30,
  minor x tick num=1,
]

\addplot+[
  no markers,
  line join=round,
] table [
  col sep=comma,
  x index=2, 
  y index=1,
] {\CurrentFilePath/hfss_alpha.csv};
\addlegendentry{Antenna}

\addplot+[
  mark=+,
  mark repeat=6,
  mark phase=3,
] table [
  x index=2, 
  y index=1,
  y expr=\thisrowno{1}*1e-9,
] {\CurrentFilePath/optmm.dat};
\addlegendentry{OPB-MNT}

\end{groupplot}

\IfStandalone{}{
  \node[anchor=south west, inner sep=0, outer sep=0, node font=\small] at (current bounding box.south west) {\textsf{(b)}};
}

\end{tikzpicture}%
  }%
  \caption{%
    (a) Unit cell of the slotted waveguide antenna. The inner cross-section corresponds to a WR90 waveguide. The slot width is 1.5875\?mm, the thickness of the metal walls is 0.5\?mm, and the other dimensions are specified in the figure.
    (b)~Dispersion diagram for the structure in (a). The results from the OPB-MNT (blue line) are compared against those extracted from a finite-length antenna simulation (red line).
  }
  \label{fig:waveguide}
\end{figure}

To validate the dispersion analysis derived from OPB-MNT, the dispersion of various classic LWA unit cells were computed and compared with data in literature, analytical solutions, and full-wave simulations.
In the following, the OPB-MNT results are also compared to those obtained using a finite-length antenna simulation. The dispersion diagram is recovered from a full-wave simulation of a finite-length antenna of a sufficiently large number of unit cells. The required number of cells is case-dependent and is chosen such that the complex wavenumber per unit of length becomes stable \cite{Deb:2026:ComparisonNumericalTechn}. The real part of the wavenumber is extracted from simulation by identifying the radiation angle with the highest power density.
The imaginary part of the wavenumber is determined from the attenuation over a length along the antenna.
The attenuation is computed by sampling and integrating the time-averaged power flow over two transverse to $\vu{x}$ surfaces separated by several whole periods.

\subsubsection{Slotted rectangular waveguide} \label{sec:slot_wg}

The first test case (shown in \figref{fig:waveguide}a) is a slotted waveguide antenna based on the dimensions of a WR90 waveguide. For this LWA, the simulation setup is modified by removing the PMC boundaries that models $y$ invariance, due to the presence of metallic waveguide walls, which are assumed to be perfect electric conductors (PECs). Above the waveguide, the boundaries remain as depicted in \figref{fig:3d_view}, meaning that the PBCs transverse to $\vu{y}$ model an array of slotted waveguides repeating along the $y$-axis. The wave ports are positioned to fit within the inner cross-section of the waveguide.

\figref{fig:waveguide}b displays the dispersion diagram, showing the real part $\beta$ (left) and imaginary part $\alpha$ (right) of $\kappa$ for this structure. The values obtained with the OPB-MNT method (blue lines), with $M=\text{1}$ mode at each wave port, are in close agreement with those from the finite-length antenna simulation (red lines) using 30 cells. The $\alpha$ curve reaches a maximum at a resonant frequency of 8.6\?GHz, where the slot length is approximately half a free-space wavelength (17.43\?mm).

\subsubsection{Slotted parallel plate waveguide}

\begin{figure}
  \centering%
  \iftoggle{pdf_instead_of_tikz}{%
    \includegraphics{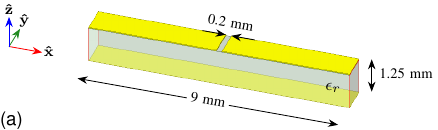}%
  }{%
    \tikzsetnextfilename{ppw}%
    \input{figure/ppw/ppw.tex}%
  }%
  \vskip5pt
  \iftoggle{pdf_instead_of_tikz}{%
    \includegraphics{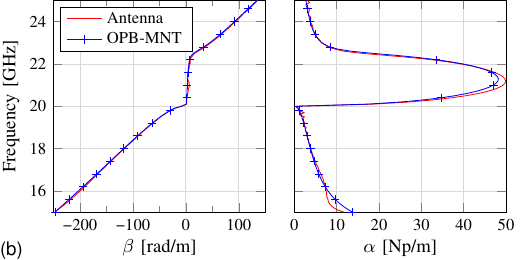}%
  }{%
    \tikzsetnextfilename{disp_ppw}%
    \begin{tikzpicture}

\begin{groupplot}[
  group2by1,
  legend pos=north west,
  ymin=15,
  ymax=25,
]

\nextgroupplot[ 
  xlabel={$\beta$ [rad/m]},
  xmin=-250,
  xmax=150,
  minor x tick num=1,
]

\addplot+[
  no markers,
  line join=round,
] table [
  x index=3, 
  y index=1,
  y expr=\thisrowno{1}*1e-9,
] {\CurrentFilePath/ref.dat};
\addlegendentry{Antenna}

\addplot+[
  mark=+,
  mark repeat=6,
] table [
  x index=3, 
  y index=1,
  y expr=\thisrowno{1}*1e-9,
] {\CurrentFilePath/optmm.dat};
\addlegendentry{OPB-MNT}

\nextgroupplot[ 
  xlabel={$\alpha$ [Np/m]},
  xmin=0,
  xmax=50,
]

\addplot+[
  no markers,
  line join=round,
] table [
  col sep=comma,
  x index=2, 
  y index=1,
] {\CurrentFilePath/hfss_alpha.csv};

\addplot+[
  mark=+,
  mark repeat=6,
] table [
  x index=2, 
  y index=1,
  y expr=\thisrowno{1}*1e-9,
] {\CurrentFilePath/optmm.dat};

\end{groupplot}

\node[anchor=south west, inner sep=0, outer sep=0, node font=\small] at (current bounding box.south west) {\textsf{(b)}};

\end{tikzpicture}%
  }%
  \caption{%
    (a)~One-dimensional periodic unit cell of the slotted parallel plate antenna. The parallel plate is filled with a dielectric with $\epsilon_r=\text{2.77}$, and its height and the slot width are specified in the figure.
    (b)~Dispersion diagram of the parallel plate antenna. The results from the OPB-MNT (blue line) are compared against those extracted from a finite-length antenna simulation (red line).
  }
  \label{fig:ppw}
\end{figure}

The second LWA considered is a PPW with periodic slots, as shown in \figref{fig:ppw}a.
\figref{fig:ppw}b reports the associated dispersion diagram, showing the real part $\beta$ (left) and the imaginary part $\alpha$ (right) of $\kappa$. The values obtained using the OPB-MNT method (blue lines), with $M=\text{1}$ mode at each wave port, are in very good agreement with those for the finite-length antenna simulation (red lines) using 30 cells. The $\alpha$ curve reaches the open stop band \cite{Jackson:2008:LWA} at the frequency of 20\?GHz, and then attains its maximum at 21.3\?GHz.

\subsubsection{Metal strip grating}

\begin{figure}
  \centering
  \iftoggle{pdf_instead_of_tikz}{%
    \includegraphics{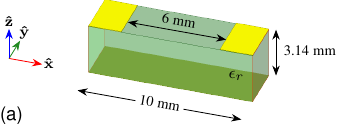}%
  }{%
    \tikzsetnextfilename{bulls-eye}%
    \input{figure/bulls-eye/bulls-eye.tex}%
  }%
  \vskip10pt%
  \iftoggle{pdf_instead_of_tikz}{%
    \includegraphics{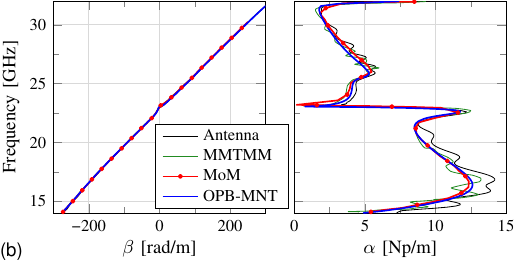}%
  }{%
    \tikzsetnextfilename{disp_bulls-eye}%
    \begin{tikzpicture}

\pgfmathsetmacro{\uclen}{10e-3}
\pgfmathsetmacro{\czero}{299792458} 

\begin{groupplot}[
  group2by1,
  minor tick num=1,
  legend style={
    anchor=south east,
    at={(1.11,0.02)}
  },
  cycle list={
    black\\
    ForestGreen,\\
    red, line width=0.8pt, mark=*, mark repeat=3, mark size=0.8pt\\
    blue,line width=0.8pt,\\
  },
  ymin=14,
  ymax=32,
]

\nextgroupplot[ 
  xlabel={$\beta$ [rad/m]},
  xmin=-300,
  xmax=300,
]

\addplot+ table [
  x index=3, 
  y expr=\thisrowno{1}*1e-9,
] {\CurrentFilePath/ref.dat};
\addlegendentry{Antenna}

\addplot+ table [
  x expr=\thisrowno{1}*pi/\uclen,
  y index=0, 
] {\CurrentFilePath/beta_mmtmm.dat};
\addlegendentry{MMTMM}

\addplot+ table [
  x expr=\thisrowno{1}*pi/\uclen,
  y index=0, 
] {\CurrentFilePath/beta_mom.dat};
\addlegendentry{MoM}

\addplot+ table [
  x index=3, 
  y expr=\thisrowno{1}*1e-9,
] {\CurrentFilePath/comsol.dat};
\addlegendentry{OPB-MNT}

\nextgroupplot[ 
  xlabel={$\alpha$ [Np/m]},
  xmin=0,
  xmax=15,
]

\addplot+ table [
  x index=2, 
  y expr=\thisrowno{1}*1e-9,
] {\CurrentFilePath/ref.dat};

\addplot+[
  smooth,
  tension=0.35,
  miter limit=2,
] table [
  x expr=\thisrowno{1}*2*pi*\thisrowno{0}*1e9/\czero,
  y index=0,
] {\CurrentFilePath/alpha_mmtmm.dat};

\addplot+ table [
  x expr=\thisrowno{1}*2*pi*\thisrowno{0}*1e9/\czero,
  y index=0,
] {\CurrentFilePath/alpha_mom.dat};

\addplot+ table [
  x index=2, 
  y index=1,
  y expr=\thisrowno{1}*1e-9,
] {\CurrentFilePath/comsol.dat};

\end{groupplot}

\IfStandalone{}{
\node[anchor=south west, inner sep=0, outer sep=0, node font=\small] at (current bounding box.south west) {\textsf{(b)}};
}

\end{tikzpicture}%
  }%
  \caption{%
  (a) One-dimensional periodic unit cell of the metal strip grating LWA printed on a grounded dielectric slab with dielectric constant $\epsilon_r=\text{2.2}$.
  (b)~Dispersion diagram of the metal strip grating LWA. The results from the OPB-MNT using a single mode ($M=\text{1}$) are compared with those extracted from a finite-length antenna simulation, the MMTMM ($M=\text{10}$) \cite{garcia2025multimodal} and the MoM \cite{comite2018radially}.
  }
  \label{fig:bull's-eye}
\end{figure}

The third unit cell considered corresponds to a periodic metal strip grating above a grounded dielectric slab.
This canonical periodic structure is the basis for simple yet effective LWA designs \cite{Honey1959Flush, Guglielmi1993Broadside}. When implemented with circular symmetry, this unit cell structure can also be used to generate Bessel beams \cite{Ettorre2012Generation, Fuscaldo:2020, Negri:2025}. In this third example, we consider the MoM dispersion analysis performed on the one-dimensional linearized structure of the configuration reported in \cite{comite2018radially} as a reference, as shown in \figref{fig:bull's-eye}a.
As in the previous examples, \figref{fig:bull's-eye}b shows the propagation constant $\beta$ (left) and the attenuation constant $\alpha$ (right). The values calculated using the OPB-MNT, employing a single mode ($M=\text{1}$) at the wave ports, are compared with those obtained using the finite-length antenna simulation using 30 cells (black line), the MoM results from \cite{comite2018radially} (red line) and the MMTMM in \cite{garcia2025multimodal} (green line). The dispersion obtained from OPB-MNT is in very good agreement with the MoM \cite{comite2018radially}, which is taken as ground truth. Although all methods provide very similar results for the real part of the complex propagation constant, the imaginary part computed with the OPB-MNT exhibits significantly fewer unwanted oscillations compared to the MMTMM \cite{garcia2025multimodal} and the finite-length antenna simulation, particularly in the lower frequency range (14 to 22\?GHz).

\subsubsection{Modulated metasurface antenna}

The fourth unit cell analyzed is a metasurface antenna.
Two different implementations of this class of LWA are studied. First, an idealized realization can be obtained by considering a modulated impedance boundary condition (IBC) over a grounded dielectric slab. Second, we studied a realizable version of the antenna where the ideal IBC is implemented by varying width metallic strips.

\begin{figure}
  \centering
  \iftoggle{pdf_instead_of_tikz}{%
    \includegraphics{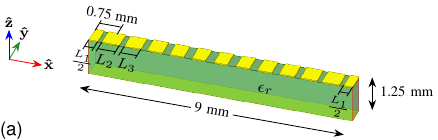}%
  }{%
    \tikzsetnextfilename{mts_metal}%
    \input{figure/mts_metal/mts_metal.tex}%
  }%
  \vskip5pt%
  \iftoggle{pdf_instead_of_tikz}{%
    \includegraphics{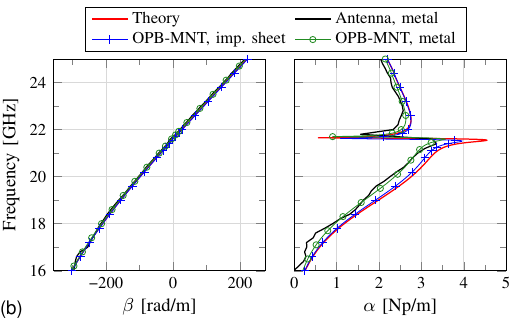}%
  }{%
    \tikzsetnextfilename{disp_mts}%
    \begin{tikzpicture}

\begin{groupplot}[
  group2by1,
  group/group name=mygroup,
  minor tick num=1,
  ymin=16,
  ymax=25,
  cycle list={
    {red, semithick},
    {black, semithick},
    {blue, mark=+},
    {ForestGreen, mark=o},
  },
]

\nextgroupplot[ 
  xlabel={$\beta$ [rad/m]},
  legend columns=2,
  legend to name=mylegend,
]

\addplot+[
  no markers,
  line join=round,
] table [
  x index=3, 
  y index=1,
  y expr=\thisrowno{1}*1e-9,
] {\CurrentFilePath/ref.dat};
\addlegendentry{Theory}

\addplot+[
  no markers,
  line join=round,
] table [
  x index=3, 
  y index=1,
  y expr=\thisrowno{1},
] {\CurrentFilePath/antenna.dat};
\addlegendentry{Antenna, metal}

\addplot+[
  mark repeat=6,
] table [
  x index=3, 
  y index=1,
  y expr=\thisrowno{1}*1e-9,
] {\CurrentFilePath/comsol.dat};
\addlegendentry{OPB-MNT, imp. sheet\;}

\addplot+[
  mark size=1.5pt,
  mark repeat=6,
  mark phase=4,
] table [
  x index=3, 
  y index=1,
  y expr=\thisrowno{1}*1e-9,
] {\CurrentFilePath/comsol_metal.dat};
\addlegendentry{OPB-MNT, metal}

\nextgroupplot[ 
  xlabel={$\alpha$ [Np/m]},
  xmin=0,
  xmax=5,
]

\addplot+[
  no markers,
  line join=round,
] table [
  x index=2, 
  y index=1,
  y expr=\thisrowno{1}*1e-9,
] {\CurrentFilePath/ref.dat};

\addplot+[
] table [
  x index=2, 
  y index=1,
  y expr=\thisrowno{1},
] {\CurrentFilePath/antenna.dat};

\addplot+[
  mark repeat=6,
] table [
  x index=2, 
  y index=1,
  y expr=\thisrowno{1}*1e-9,
] {\CurrentFilePath/comsol.dat};

\addplot+[
  mark size=1.5pt,
  mark repeat=6,
] table [
  x index=2, 
  y index=1,
  y expr=\thisrowno{1}*1e-9,
] {\CurrentFilePath/comsol_metal.dat};

\end{groupplot}

\node[anchor=south] at ($(mygroup c1r1.north)!0.5!(mygroup c2r1.north)-(2pt,0)$) {\ref{mylegend}};

\IfStandalone{}{
\node[anchor=south west, inner sep=0, outer sep=0, node font=\small] at (current bounding box.south west) {\textsf{(b)}};
}

\end{tikzpicture}%
  }%
  \caption{%
    (a) One-dimensional periodic unit cell of the modulated metasurface antenna implemented with metallic strips printed on a grounded dielectric slab with $\epsilon_r=\text{6.15}$.
    (b)~Dispersion diagram of the metasurface antenna. The OPB-MNT is applied to two structures: the ideal version of the antenna (modeled with a transparent impedance sheet) and its implementation using metal strips. For the former, a reference solution based on Oliner's theory is computed (red line), while for the latter, the reference values are obtained from a finite-length antenna simulation (black line) based on the unit cell shown in \figref{fig:mts}a.
  }
  \label{fig:mts}
\end{figure}

In the first case, the considered IBC corresponds to a penetrable or transparent impedance sheet condition \cite{Kuester2003Averaged, Patel2011PrintedLeaky, Holloway2012Overview, Patel2013Modeling, Glez:2015, Mencagli:2016}, as opposed to an impenetrable one \cite{Glisson1992Electromagnetic, Bleszynski1993Surface, Francavilla2015Numerical}. The sinusoidal modulation of the transparent sheet reactance is defined as
\begin{equation}
    X_{\rms}(x)=X_{\rm av}\mathopen{}\bqty{1+M\cos(\mfrac{2\pi}{d}x)}, 
\end{equation}
with $X_{\rm av}={}$\textminus548.5\?\textohm\ and $M=\text{0.4}$.
The reference solution for this ideal problem has been computed using Oliner's method \cite{oliner_guided_1959}, as adapted in \cite{Caminita} for transparent impedance sheets. This solution, labeled as theory in \figref{fig:mts}b, is used as a reference for the ideal case.
To assess the performance of the OPB-MNT method, a metasurface unit cell implemented using a transparent impedance sheet was considered. The unit cell is discretized into 12 equal-sized sections, each with a constant sheet reactance, whose values are listed in Table~\ref{tab:mts}.
The results obtained with the OPB-MNT method for $M=\text{1}$ (blue curve with + markers) are compared with Oliner's theory (red curve) in \figref{fig:mts}b. An excellent agreement is observed between the two results.

In the second case, the ideal modulated IBC is implemented using a patterned metallic cladding \cite{Patel2013Controlling, Patel2011PrintedLeaky}, as shown in \figref{fig:mts}a. The center-to-center distance between the metal strips is constant and equal to 0.75\?mm, while the metal strip width varies from section to section and is chosen to implement the corresponding sheet reactance. The specific values, listed in Table~\ref{tab:mts}, were found by building a look-up table that maps the metal strip size to the equivalent sheet reactance by performing a set of eigenmode simulations using \textsl{HFSS} \cite{HFSS}.
A reference solution for this structure with 20 cells was also computed using the finite-length antenna simulation.
The finite-length antenna simulation results (black curves) are compared with those obtained by analyzing the unit cell using the OPB-MNT with $M=\text{1}$ (green curve with $\circ$ markers) in \figref{fig:mts}b. Once again, very good agreement is observed between the two approaches.

\begin{table}
  \centering
  \caption{Section Parameters of the Metasurface Unit Cell}
  \label{tab:mts}
  \vskip-\medskipamount
  \begin{tblr}{
    width=252pt,
    colspec={lr|X[c]X[c]X[c]X[c]X[c]X[c]},
    colsep=2.9pt,
    rows={ht=0.9\baselineskip},
    column{1}={cmd={\smash[b]}, rightsep=0pt, font=\bfseries\boldmath},
    column{2}={cmd={\smash[b]}, leftsep=0pt, preto=\space},
    cell{2}{3-Z}={cmd=\hyphentominus},
    cell{5}{3-Z}={cmd=\hyphentominus},
    stretch=-1,
    rulesep=1ex,
  }
    \hline
    Section $n$                     &             & 1      & 2      & 3      & 4      & 5      & 6      \\
    \hline[dotted, lr, endpos]
    Reactance                       & [\textohm]  & -341.1 & -371.6 & -454.8 & -568.5 & -682.3 & -765.5 \\
    \hline[dotted, lr, endpos]
    Strip width $L_n$ & [\textmu m] & 617    & 602    & 565    & 521    & 485    & 463    \\
    \hline
    \hline
    Section $n$                     &             & 7      & 8      & 9      & 10     & 11     & 12     \\
    \hline[dotted, lr, endpos]
    Reactance                       & [\textohm]  & -796.0 & -765.5 & -682.3 & -568.5 & -454.8 & -371.6 \\
    \hline[dotted, lr, endpos]
    Strip width $L_n$ & [\textmu m] & 455    & 463    & 485    & 521    & 565    & 602    \\
    \hline
  \end{tblr}
\end{table}

When comparing the two implementations, the real part ($\beta$) of the wavenumber, shown in \figref{fig:mts}b, exhibits excellent agreement across all analysis methods. Both metasurface implementations analyzed with the OPB-MNT---the impedance sheet model (blue curve with + markers), and the metallic strips implementation (green curve with $\circ$ markers)---closely follow the theoretical prediction (red curve). They also match the results obtained with the finite-length antenna simulation (black curve).
Regarding the imaginary part ($\alpha$) of $\kappa$, the impedance sheet implementation also closely follows the theory. In contrast, the metallic strips implementation exhibits slightly lower values of $\alpha$ over the analyzed frequency range, which is consistent with the value obtained with the finite-length antenna simulation.

\subsubsection{Periodic Gap-Coupled Microstrip Line}

\begin{figure}
  \centering
  \includegraphics[width=0.55\linewidth]{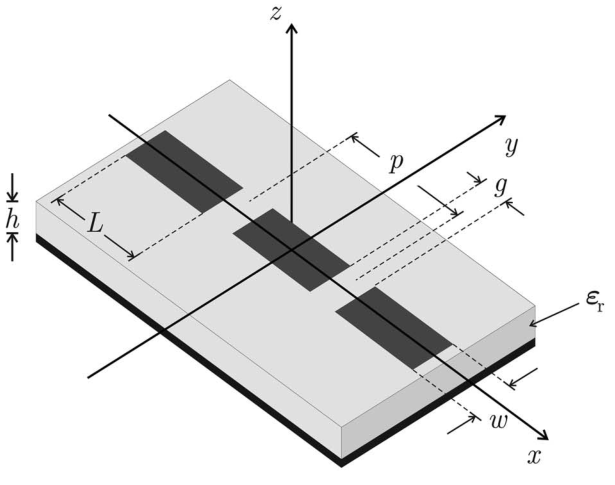}%
  \includegraphics[width=0.45\linewidth]{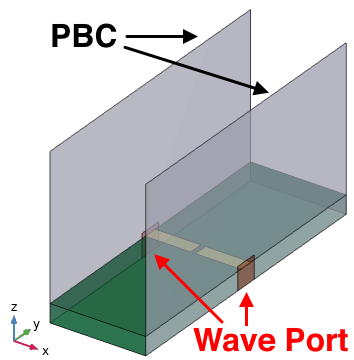}\\
  \makebox[0.55\linewidth][c]{\sffamily\small(a)}%
  \makebox[0.45\linewidth][c]{\sffamily\small(b)}\\[\medskipamount * 13/10]
  \includegraphics[width=0.60\linewidth,trim=0 25 450 260,clip]{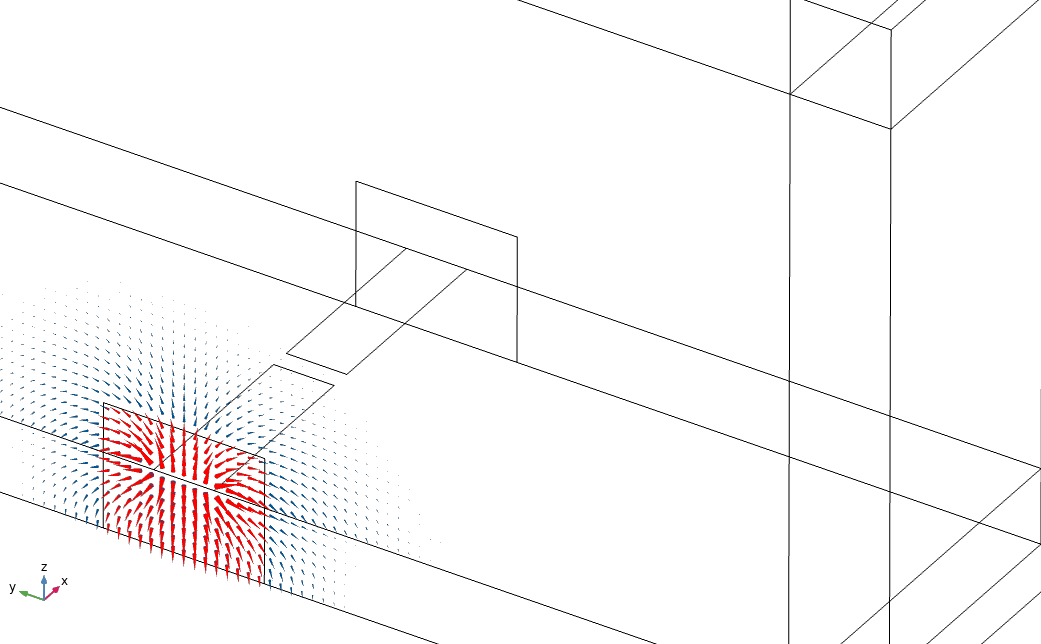}\\[-1.7ex]
  {\sffamily\small(c)}\par
  \vspace{\medskipamount * 7/10}
  \begin{tikzpicture}

\pgfkeys{/pgf/fpu}
\pgfmathsetmacro{\uclen}{4e-3}
\pgfmathsetmacro{\czero}{299792458} 
\pgfmathsetmacro{\kAtFourtyGHz}{2*pi*40e9/\czero} 
\pgfkeys{/pgf/fpu=false}

\begin{groupplot}[
  group2by1,
  minor tick num=1,
  legend pos = north east,
  cycle list={
    black, line width=0.5pt, mark=+, mark repeat=3, mark phase=2\\
    red, line width=0.5pt, mark=*, mark repeat=3, mark size=0.8pt\\
    blue,line width=0.8pt,\\
  },
  ymin=13,
  ymax=40,
]

\nextgroupplot[ 
  xlabel={$\beta p$ [rad]},
  xmin=0,
  xmax=3.1,
  x tick label style={name=xtick},
  scaled x ticks=manual:{$\cdot\pi$}{},
  x tick scale label style={
    at={(xticklabel cs:1)},
    anchor=north east,
  },
]

\addplot[forget plot, densely dotted, line width=0.5pt] 
  coordinates {
    (0,0)
    ({\kAtFourtyGHz*\uclen/pi},40)
  };

\addplot[forget plot, densely dotted, line width=0.5pt, shift={(axis direction cs:2,0)}] 
  coordinates {
    ({-\kAtFourtyGHz*\uclen/pi},40) 
    (0,0) 
    ({\kAtFourtyGHz*\uclen/pi},40)
  };

\addplot+ table [
  x=beta, 
  y=freq,
  col sep=comma,
] {\CurrentFilePath/disp_microstrip_hfss.csv};

\addplot+ table [
  x=beta, 
  y=freq_beta,
  col sep=comma,
] {\CurrentFilePath/disp_microstrip_reference_beta.csv};

\addplot+ table [
  x=beta, 
  y=freq,
  col sep=comma,
] {\CurrentFilePath/disp_microstrip_comsol.csv};

\nextgroupplot[ 
  xlabel={$\alpha / k_0$ [Np/rad]},
  xmin=0,
  xmax=1.25,
]

\addplot+ table [
  x index=2,
  y index=1,
  col sep=comma,
] {\CurrentFilePath/disp_microstrip_hfssA.csv};
\addlegendentry{Antenna}

\addplot+ table [
  x=alpha, 
  y=freq_alpha,
  col sep=comma,
] {\CurrentFilePath/disp_microstrip_reference_alpha.csv};
\addlegendentry{MPIE}

\addplot+ table [
  x=alpha, 
  y=freq,
  col sep=comma,
] {\CurrentFilePath/disp_microstrip_comsol.csv};
\addlegendentry{OPB-MNT}

\end{groupplot}

\IfStandalone{}{
\node[anchor=south west, inner sep=0, outer sep=0, node font=\small\sffamily] 
  at (current bounding box.south west) {(d)};
}

\end{tikzpicture}%
  \caption{%
    (a)~Geometry of the gap-coupled microstrip line LWA \cite{Valerio:2015:EfficientComp1DLWA} printed on a substrate with $\epsilon_{\mathrm{r}}={}$10.6. Dimensions: $p={}$0.4\?mm, $L={}$3.8\?mm, $w={}$0.6\?mm, $h={}$0.762\?mm.
    (b)~COMSOL model of the unit cell used for the dispersion analysis. The extents of the wave ports and PBCs are indicated. The other boundaries normal to $\vu{y}$ or $\vu{z}$ are terminated with Perfectly Matched Layers (PMLs).
    (c)~Result of the preliminary eigenmode analysis. The quasi-TEM mode fields are computed over the entire cross-section, but only the fields highlighted in red are used to specify the user-defined wave ports for the dispersion analysis.
    (d)~Dispersion diagram of the gap-coupled microstrip line. The results obtained using OPB-MNT are compared with those extracted from a finite-length antenna simulation and with those reported in \cite{Valerio:2015:EfficientComp1DLWA}, obtained by solving a MPIE. The dotted lines are the light lines $\omega=|k|c$, and their periodic replicas.
  }
  \label{fig:microstrip}
\end{figure}

The fifth unit cell considered corresponds to a microstrip line LWA. We consider here the design analyzed in \cite{Valerio:2015:EfficientComp1DLWA}, which is reproduced for convenience in \figref{fig:microstrip}a. Gaps in the microstrip line provide the necessary periodic discontinuities. 

The simulation setup is depicted in \figref{fig:microstrip}b. It shows the location and extent of the wave ports and PBCs. In this antenna, the input port is not confined by two metallic and two PMC boundaries. Indeed, the microstrip line supports a guided quasi-TEM mode and there is no clear distinction between a guided and a radiation region.  In this configuration, part of the outer boundary of the port is in direct contact with the surrounding PBC, as shown in \figref{fig:microstrip}b, unlike in the previous examples, where they are separated by metallic plates.

This new simulation setup prevents the standard wave port eigenmode analysis from being performed directly. Instead, because the PBC and the wave ports are in direct contact, the wave ports [shown as small brown areas in \figref{fig:microstrip}b] are defined as user-defined with a prescribed field distribution. Before defining these user-defined ports, a preliminary eigenmode analysis is performed on the port plane, with the PBCs removed and the port analysis area extended to cover also the PBC [the gray surface in \figref{fig:microstrip}b].
The solution obtained from the preliminary eigenmode analysis over the larger cross-section, marked in gray in \figref{fig:microstrip}b, is then used to define the smaller user-defined wave ports. This is done by truncating the field distribution obtained over the larger cross-section to the extent of the smaller brown areas shown in \figref{fig:microstrip}b. Only the field lines highlighted in red in \figref{fig:microstrip}c are retained, and these define the wave port mode used in the subsequent dispersion analysis.

After this preliminary wave port mode analysis, the standard OPB-MNT dispersion analysis described in Section \ref{sec:dispersion} is conducted. The results are presented in \figref{fig:microstrip}d. The values calculated using the OPB-MNT, with a single quasi-TEM mode specified at the wave ports, are compared with those obtained from a finite-length antenna simulation using 20 unit cells (black line), and the mixed-potential integral equation (MPIE) results reported in  \cite{Valerio:2015:EfficientComp1DLWA} (red line). The dispersion results obtained using OPB-MNT are in good agreement with the MPIE results in both the bounded region (13--22.9\?GHz) and the radiation region (22.9--40\?GHz). In the radiation region, the OPB-MNT results agree particularly well with the finite length antenna simulation, except within  the open stop band (26.7--32\?GHz), where the finite-length antenna simulation results become inaccurate.

\subsubsection{Dual-Mode Waveguide}\label{Sec:dual-mode}

The final LWA analyzed is a dual-mode waveguide presented in \cite{Dorle:2021:CPLWADualModeWG}. The rectangular waveguide supports the TE\textsubscript{10} and the TE\textsubscript{01} modes, as depicted in \figref{fig:dualmode_wg}a. The interaction of these two modes with the periodic cross-shaped slots can produce circularly polarized radiation. As in the case of the slotted rectangular waveguide (\textsection\?\ref{sec:slot_wg}), the wave ports fit within the inner section of the waveguide, while the remaining boundaries above it are configured as depicted in \figref{fig:3d_view}. The wave ports are defined to excite $M={}$2 modes: the TE\textsubscript{10} and the TE\textsubscript{01}.

To retrieve the dispersion of the periodic antenna, the OPB-MNT method is applied twice, once for each mode, in order to converge separately to the corresponding solution. This requires discriminating between the eigenvalues at each iteration so that the one corresponding to the same mode is selected at every iteration of the algorithm. This can be achieved by checking that the selected eigenvalue corresponds to the eigenvector defining the mode of interest across successive iterations.
The results are presented in \figref{fig:dualmode_wg}b. The values calculated using the OPB-MNT (solid lines), are compared with those obtained from a finite-length antenna simulation using 20 unit cells (dashed lines). The dispersion curves obtained using OPB-MNT are in good agreement with those from the finite-length antenna simulation.

\begingroup
\begin{figure}
  \centering
  {\sffamily\small(a)\hspace{1.5em}}%
  \includegraphics[width=0.37\linewidth]{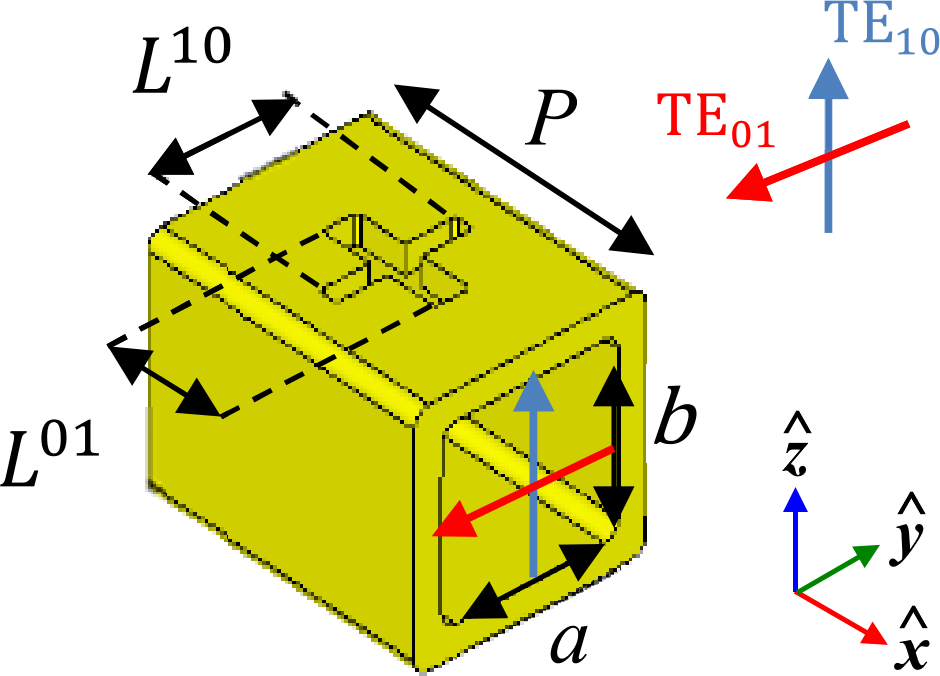}%
  \vspace{\medskipamount}
  \tikzsetnextfilename{disp_dualmode_wg}%
  \begin{tikzpicture}

\begin{groupplot}[
  group2by1,
  minor tick num=1,
  legend pos = south east,
  legend style={
    anchor=south east,
    at={(1.00,0.00)},
    shift={(135:2pt)}
  },
  cycle list={
    red!50!yellow!90!black,  densely dashed  \\
    blue!50!cyan!90!black, densely dashed  \\
    red\\
    blue\\
  },
  every axis plot/.append style={line width=0.6pt},
  ymin=17,
  ymax=23,
  ytick distance=1,
]

\nextgroupplot[ 
  xlabel={$\beta P$ [rad]},
  xmin=0,
  xmax=1,
  scaled x ticks=manual:{$\cdot\pi$}{},
  x tick scale label style={
    at={(xticklabel cs:1)},
    anchor=north east,
  },
]



\addplot+ table [
  x=beta_TE10, 
  y=freq_beta,
  col sep=comma,
] {./\CurrentFilePath/disp_wg_dualmode_hfss_beta.csv};

\addplot+ table [
  x=beta_TE01, 
  y=freq_beta,
  col sep=comma,
] {./\CurrentFilePath/disp_wg_dualmode_hfss_beta.csv};

\addplot+ table [
  x=beta_TE10, 
  y=freq,
  col sep=comma,
] {./\CurrentFilePath/disp_wg_dualmode_comsol.csv};

\addplot+ table [
  x=beta_TE01, 
  y=freq,
  col sep=comma,
] {./\CurrentFilePath/disp_wg_dualmode_comsol.csv};

\nextgroupplot[ 
  xlabel={$\alpha / k_0$ [cNp/rad]},
  restrict y to domain=16.8:23,
  xmin=0,
  xmax=12,
]

\addplot+ table [
  x=alpha_ov_k_TE10, 
  y=freq_alpha,
  col sep=comma,
] {./\CurrentFilePath/disp_wg_dualmode_hfss_alpha.csv};
\addlegendentry{Antenna {TE\textsubscript{10}}}

\addplot+ table [
  x=alpha_ov_k_TE01, 
  y=freq_alpha,
  col sep=comma,
] {./\CurrentFilePath/disp_wg_dualmode_hfss_alpha.csv};
\addlegendentry{Antenna {TE\textsubscript{01}}}

\addplot+ table [
  x=alpha_ov_k_TE10, 
  y=freq,
  col sep=comma,
] {./\CurrentFilePath/disp_wg_dualmode_comsol.csv};
\addlegendentry{OPB-MNT {TE\textsubscript{10}}}

\addplot+ table [
  x=alpha_ov_k_TE01, 
  y=freq,
  col sep=comma,
] {./\CurrentFilePath/disp_wg_dualmode_comsol.csv};
\addlegendentry{OPB-MNT {TE\textsubscript{01}}}

\end{groupplot}

\IfStandalone{}{
\node[anchor=south west, inner sep=0, outer sep=0, node font=\small\sffamily] 
  at (current bounding box.south west) {(b)};
}

\end{tikzpicture}%
  \caption{%
    (a)~Geometry of the dual-mode waveguide unit cell \cite{Dorle:2021:CPLWADualModeWG}. The dimensions are: $P={}$7\?mm, $a={}$9\?mm, $b={}$8.8\?mm, $L^{10}={}$7.3\?mm, $L^{01}={}$6.2\?mm, and the metal thickness is 1.4\?mm. The slot widths are 1\?mm in the $x$-direction and 1.76\?mm in the $y$-direction. 
    (b)~Dispersion diagram of the dual-mode waveguide. The results obtained using OPB-MNT for the TE\textsubscript{10} and TE\textsubscript{01} modes are compared with those extracted from a finite-length antenna simulation.
  }
  \label{fig:dualmode_wg}
\end{figure}
\endgroup

\subsubsection{Summary}

In Table~\ref{tab:iteration}, we have provided information on the number of iterations and the computation time required per frequency point to reach convergence for all the LWAs discussed above. These statistics were recorded with a convergence threshold $\epsilon_\kappa$ set to 0.005\,$k_0$. Despite the small value of $\epsilon_\kappa$, the average number of iterations remains fairly low for all cases over the frequency range of interest.  On the computer used (CPU Intel Xeon 3.30\?GHz, 256\?GB of RAM), the mean computation time required to achieve convergence for a single frequency point is around 2\?min for the tree-dimensional unit cells (i.e.\@ the slotted waveguide, the gap-coupled microstrip line, and the dual-mode waveguide), and around 1\?min for the two-dimensional (planar) unit cells.  It should be noted that, within an iteration at a given frequency, nearly all the computation time is spent on the full-wave simulation in the dedicated commercial solver (over 99.5\% of the iteration time). In contrast, the post-processing stage of the OPB-MNT method requires only about 25\?ms in the case of a single wave port mode. This post-processing computational time primarily depends on the number of wave port modes, since it determines the size of the matrix \eqref{eq:matrix-eigen} and, thus, the time required to solve the eigenvalue problem.

\begingroup
\begin{table}
  \sisetup{number-mode=match}%
  \caption{Number of Iterations and Convergence Time\\for a Single Frequency Point}
  \label{tab:iteration}
  \centering
  \vspace{-\medskipamount}
  \begin{lrbox}{\boxA}
  \begin{booktabs}{
    colspec={lccS[table-format=2]cS[table-format=2.2]},
    cell{1}{1} = {r=2}{},
    cell{1}{2} = {c=3}{},
    cell{1}{5} = {c=2}{},
    row{1-2} = {valign=m, guard},
    hspan=even,
  }
  \toprule
    {Leaky-Wave Antenna\\\emph{Variation}}   & {Number of Iterations\\OPB-MNT} &&& Mean Time [min] \\
  \cmidrule[lr]{2-4}
  \cmidrule[lr]{5-6}
                                             & mean & $\sigma$ & max          & OPB-MNT & FLAS*\\
  \midrule                                                                      
    Slotted Waveguide                        & 2.01 & 0.19     &  3           & 1.67    & 48.00  \\ 
    Slotted PPW                              & 2.86 & 0.55     &  4           & 0.76    &  1.42  \\ 
    Metal Strip Grating                      & 3.91 & 0.28     &  4           & 1.05    & 13.21  \\ 
    MTS \emph{IBC Sheet}                     & 4.29 & 0.66     &  7           & 1.16    &  4.90  \\ 
    MTS \emph{Metal Strips}    & 4.47 & 1.02     & 12           & 1.23    & 25.67  \\ 
    Gap Microstrip Line                      & 4.12 & 3.72     & 14           & 2.33    & 65.60  \\ 
    Dual-Mode WG \emph{TE\textsubscript{10}} & 1.88 & 0.49     &  2           & 2.03    & 35.52  \\ 
    Dual-Mode WG \emph{TE\textsubscript{01}} & 1.89 & 0.50     &  3           & 2.06    & 35.52  \\ 
  \bottomrule
  \end{booktabs}%
  \end{lrbox}
  \parbox{\wd\boxA}{%
    \scriptsize
    \usebox{\boxA}\\[0.5ex]
    *Finite-Length Antenna Simulation
  }
\end{table}
\endgroup

\section{Driven Problem Using OPB-MNT} \label{sec:receiving}

In this section, we describe the second method derived within the OPB-MNT framework. Here, the response of the periodic LWA is considered under the illumination of an exterior radiation. It is assumed that an incoming plane wave is incident on each unit cell with constant power density over the unit cell length. The plane wave is described by its real wave vector $\vb{k} = k\,\vu{k} = k_x\vu{x} + k_z\vu{z}$ and incident angle $\theta$ with respect to $\vu{z}$ (see \figref{fig:3d_view}) with $k=\flatfrac{2\pi f}{c}$ and such that $k_x = \beta+\flatfrac{2\pi n}{d}$, ($n \in \bbZ$), and $\beta = k\sin\theta$.
Because there is no attenuation ($\alpha=\text{0}$) from one end of the unit cell to the next, the driven problem needs to be treated differently. In this section, we will solve for the circuit parameters (voltage, current, power, and impedance) at the Floquet and wave ports. We will consider the driven problem for the metal strip grating and metasurface antenna unit cells, and compare them to the values we obtained in the dispersion problem.

The total port voltages and currents $\sbar{V}_2$, $\sbar{I}_2$ and $\sbar{I}_3$ are computed by solving the matrix equation in \eqref{eq:matrix-driven} for a given voltage excitation $\sbar{V}_{\rm s}$ and incident angle $\theta$.
To this end, a full-wave simulation of the setup shown in \figref{fig:3d_view} is performed for the specific incident angle $\theta$. This provides the Z\=/parameters and $\kappa=k\sin\theta$, necessary to build the matrix in \eqref{eq:matrix-driven}.
Once the voltages and currents at port 2 are found using \eqref{eq:matrix-driven}, $\sbar{V}_1$ and $\sbar{I}_1$ are computed using \eqref{eq:VIejkd} while $\sbar{V}_3$ is computed using \eqref{eq:VI-floquet}. The power at ports 1 or 2 (denoted as $i$) entering the unit cell for mode $n$ is
\begin{equation}
  P_i^{(n)} = V_i^{(n)} {I_i^{(n)}}^*.
\end{equation}
We define the modal \emph{periodic driven impedance} in the negative and positive directions as:
\begin{align}
  Z_{-}^{(n)} &= -\frac{V_1^{(n)}}{I_1^{(n)}}, &
  Z_{+}^{(n)} &= -\frac{V_2^{(n)}}{I_2^{(n)}}.
\end{align}
Terminating port 1 with $Z_-$ or port 2 with $Z_+$ eliminates reflections for the driven unit cell under the excitation of a plane wave, in a manner similar to that in which the Bloch impedance eliminates reflections in an eigenvalue (modal) problem \cite{Pozar:2011:MicroEngin}.

Since the full S-parameter network is recovered during the driven problem simulation, it can be used for subsequent circuit analysis. For example, the response of a unit cell connected to different feeds or load impedances can be simulated to perform a power budget analysis for a given angle of incidence and frequency. In the circuit solver, terminating a unit cell wave port with the periodic driven impedance allows the periodicity in one direction to be modeled.
Another possibility is to simulate a cascade of different unit cell configurations connected in series to represent a modulation taper in a metasurface antenna. These subsequent circuit simulations complete and extend the driven analysis of the structure.

\subsection{Numerical Results}
Here, we present the real power and the periodic driven impedance resulting from a TM polarized plane wave excitation.
The transverse time-averaged power delivered by the plane wave through the Floquet port is $P(\theta)=P_0\cos\theta$, where $P_0$ is set to 1\?W.
On the other hand, the time-averaged power delivered by the source voltage to the Floquet port is given by $P(\theta) = \abs{V_{\rms}}^2 \!/ ({8Z_{\mathrm{TM}}(\theta)})$ \cite{Pozar:2011:MicroEngin}. Therefore, the source voltage is $V_{\rms}(\theta) = \sqrt{8Z_0 P_0} \cos \theta$, with $Z_0$ equal to the characteristic impedance of free space.
The wave ports 1 and 2 are described by a single TEM port mode.
The Floquet port is defined to include 6 Floquet modes. The only traveling modes are the TM\textsubscript{0} and TE\textsubscript{0}. The higher order Floquet modes are introduced to reduce reflections at the Floquet port boundary and improve the accuracy of the results.

\subsubsection{Metal strip grating}
First, we consider the metal strip grating antenna. The configuration in \cite{comite2018radially}, as shown in \figref{fig:bull's-eye}a is taken as a reference. After solving for $V_1$ and $I_1$ using \eqref{eq:matrix-driven}, we compute the power at port 1 for incident angles ranging from \textminus90\textdegree{} to 90\textdegree{} and for frequencies ranging from 14\?GHz to 32\?GHz. We are interested in the real part of $P_1$, since the real power at port 1 denotes the time-averaged power guided from one unit cell to the next. The real part of $P_1$ is shown in \figref{fig:power2d_bull}.
Regions with high real power identify those frequencies and incident angles that allow maximum power conversion from the incident plane wave to the guided wave. As an example, the high positive real power region starting from (14\?GHz, \textminus68\textdegree) and ending at (32\?GHz, 28\textdegree) corresponds to the main radiating harmonic, computed earlier in Section~\ref{sec:dispersion} and shown in \figref{fig:bull's-eye}b.

\begingroup
\begin{figure}
  \centering
  \includegraphics{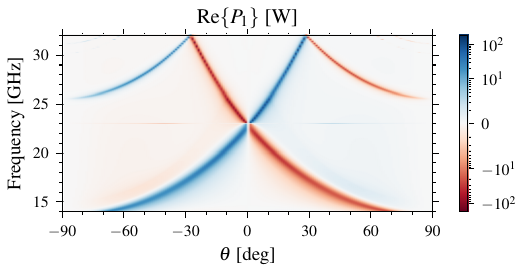}%
  \caption{Real power at port 1 of the metal strip grating unit cell as a function of frequency and incident angle for a TM polarized incoming plane wave with a power of 1\?W.}
  \label{fig:power2d_bull}
\end{figure}

\begin{figure}
  \centering%
  \iftoggle{pdf_instead_of_tikz}{%
    \includegraphics{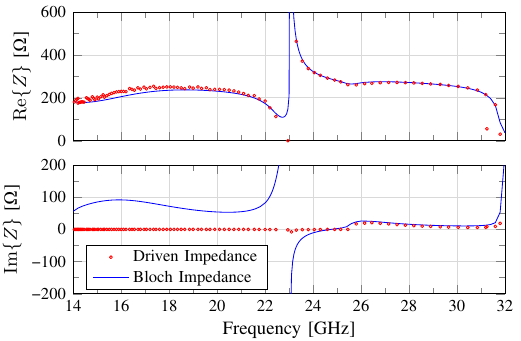}%
  }{%
    \tikzsetnextfilename{Z2_bull}%
    \begin{tikzpicture}

\begin{groupplot}[
  group1by2,
  width=208pt,
  height=62pt,
  xmin=14,xmax=32,
  legend pos=south west,
  minor tick num=1,
]

\nextgroupplot[
  ymin=0,
  ymax=600,
  ylabel={$\Real{Z}$ [\textohm]},
]

\addplot+[
  mark=o,
  mark size={0.7pt},
  only marks,
] table [
  x index=1, 
  y index=2,
  col sep=comma,
] {\CurrentFilePath/driven_bulls-eye_sweep_6modes_Z2.csv};

\addplot+[
] table [
  x index=1, 
  y index=2,
] {\CurrentFilePath/eigen_bulls_eye_Z2.dat};

\nextgroupplot[ 
  ymin=-200, ymax=200,
  restrict y to domain=-350:350,
  ylabel={$\Imag{Z}$ [\textohm]},
]


\addplot+[
  mark=o,
  mark size={0.7pt},
  only marks,
] table [
  x index=1, 
  y index=3,
  col sep=comma,
] {\CurrentFilePath/driven_bulls-eye_sweep_6modes_Z2.csv};
\addlegendentry{Driven Impedance}

\addplot+[
  mark size={0.4pt},
] table [
  x index=1, 
  y index=3,
] {\CurrentFilePath/eigen_bulls_eye_Z2.dat};
\addlegendentry{Bloch Impedance}

\end{groupplot}

\end{tikzpicture}%
  }%
  \caption{Comparison between the periodic driven impedance (driven problem) and the Bloch impedance (eigen problem) for the metal strip grating unit cell.}
  \label{fig:bloch_bull}
\end{figure}
\endgroup

We now compute the positive periodic driven impedance $Z_+$ for this set of $(f,\theta)$ coordinates that belongs to the main radiating harmonic and compare it with the Bloch impedance values obtained from the currents and voltages calculated using the dispersion analysis method \eqref{eq:matrix-eigen}. Comparing impedances, rather than voltages or currents, is necessary as it removes the dependence on $V_{\rms}$ in the driven problem, thereby enabling a fair comparison. The computed impedances are shown in \figref{fig:bloch_bull}.
Although the general trends are similar in the two cases, indicating that the driven and eigen problems are related, several differences can be observed. First, the imaginary part of $Z_+$ varies with frequency in the eigenvalue problem, while it remains zero for the driven problem. Second, at lower frequencies (14 to 22\?GHz), $\Real{Z_+}$ is slightly higher in the driven problem than in the eigen problem.

The differences between the driven and eigen problems are due to different boundary conditions. Specifically, the phase delay $\kappa d$ defining the PBCs and the one computed at the wave ports (through the post-processing computations) are different for the two problems.
In the driven problem, the wavenumber is strictly real, $\kappa=\beta$, while in the eigen problem it is complex, $\kappa=\beta-j\alpha$.

\subsubsection{Modulated metasurface antenna}
The second periodic LWA analyzed as a driven problem is the metasurface antenna implemented with metallic strips, shown in \figref{fig:mts}a. As in the case of the metal strip grating unit cell considered earlier, the real power at port 1 is computed using \eqref{eq:matrix-driven} for a TM polarized plane wave. The resulting real power is presented in \figref{fig:power2d_mts-metal}.
\begingroup
\begin{figure}
  \centering
  \includegraphics{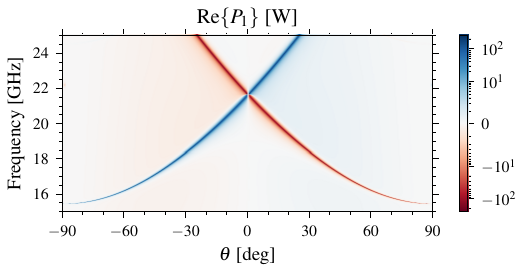}
  \caption{Real power at port 1 of the metasurface unit cell implemented with metal strips as a function of frequency and incident angle for a TM polarized incoming plane wave with a power of 1\?W.}
  \label{fig:power2d_mts-metal}
\end{figure}
The real power at port 1 resembles that obtained for the strip grating unit cell, with most of the plot exhibiting low power and only thin curves showing high real-power magnitudes. The region of high positive real power, starting at (15.5\?GHz, \textminus90°) and ending at (25\?GHz, 25°), corresponds to the radiating spatial harmonic previously identified using the iterative eigenvalue-based method in Section~\ref{sec:dispersion}.

\begin{figure}
  \centering%
  \iftoggle{pdf_instead_of_tikz}{%
    \includegraphics{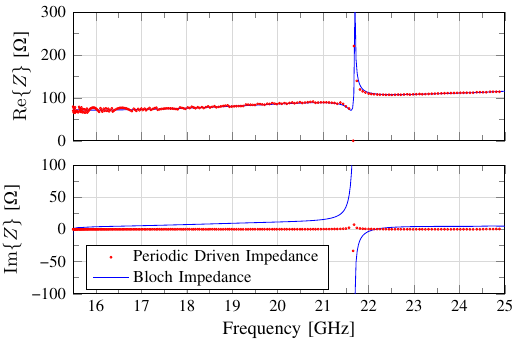}%
  }{%
    \tikzsetnextfilename{Z2_mts-metal}%
    \begin{tikzpicture}

\begin{groupplot}[
  group1by2,
  width=208pt,
  height=62pt,
  xmin=15.5,xmax=25,
  minor x tick num=1,
  minor y tick num=1,
  legend pos=south west,
]

\nextgroupplot[
  ymin=0, ymax=300,
  ylabel={$\Real{Z}$ [\textohm]},
]

\addplot+[
  mark=o,
  mark size={0.5pt},
  only marks,
] table [
  x index=1, 
  y index=2,
  col sep=comma,
] {\CurrentFilePath/driven_mts_metal_Z2.csv};

\addplot+[
] table [
  x index=1, 
  y index=2,
] {\CurrentFilePath/eigen_mts-metal_Z2.dat};

\nextgroupplot[ 
  ymin=-100, ymax=100,
  restrict y to domain=-150:150,
  ylabel={$\Imag{Z}$ [\textohm]},
]


\addplot+[
  mark=o,
  mark size={0.5pt},
  only marks,
] table [
  x index=1, 
  y index=3,
  col sep=comma,
] {\CurrentFilePath/driven_mts_metal_Z2.csv};
\addlegendentry{Periodic Driven Impedance}

\addplot+[
  mark size={0.4pt},
] table [
  x index=1, 
  y index=3,
] {\CurrentFilePath/eigen_mts-metal_Z2.dat};
\addlegendentry{Bloch Impedance}

\end{groupplot}

\end{tikzpicture}%
  }%
  \caption{Comparison between the periodic driven impedance and the Bloch impedance for the metasurface antenna unit cell implemented with metallic strips.}
  \label{fig:bloch_mts-metal}
\end{figure}

As before, the positive periodic driven impedance is computed and compared with the Bloch impedance of the eigen problem in \figref{fig:bloch_mts-metal}.
Here again, $\Imag{Z_+}$ remains zero over the entire frequency range for the driven problem, while it varies with frequency for the eigen problem.
Examining the real part of $Z_+$, the two curves follow each other very closely, exhibiting the same open stop band at 21.7\?GHz. The aliasing observed at lower frequencies is due to the low frequency sampling rate in \figref{fig:power2d_mts-metal}.

\section{Conclusion}    \label{sec:conclusion}

This paper presented a new framework for analyzing periodic LWA, referred to as the Multi-modal Network Theory with Open Periodic Boundary (OPB-MNT). It is a hybrid framework based around the full-wave simulation of a single unit cell using a commercial solver in a partially periodic environment, coupled with a post-processing step. 
The OPB-MNT offers several advantages over previous methods. First, by relying on PBCs to account for inter-cell coupling, the simulation domain is reduced to a single unit cell. Secondly, the hybrid nature of the framework eliminates the need to implement a custom solver. 
Two different LWA analysis methods are derived is framework.

The first allows for the eigen-analysis of a periodic LWA. The problem is solved using an iterative method which enables accurate recovery of the dispersion characteristics of periodic LWAs. The method is validated by analyzing several classic periodic LWAs from literature, and comparing the dispersion diagram obtained with OPB-MNT with other analytical and numerical methods. Compared to the related MMTMM, the OPB-MNT requires fewer modes for accurate analysis. It is also independent of the height of the open-region in the simulation. This eliminates the need for a convergence analysis of the port heights, thereby simplifying the procedure. These benefits come at the cost of requiring full-wave simulations for each iterative step of the frequency of interest. Nevertheless, we showed that the algorithm converged fairly rapidly despite imposing a strict convergence criterion.

The second method allows for the analysis of the driven problem, where a plane-wave is incident to the LWA. The response of the LWA in a periodic environment is computed by solving a simple matrix equation after performing a full-wave simulation. This method is validated by analyzing LWAs with two different unit cell designs and by comparing the results with those obtained from eigen-analysis.
The comparisons showed that the eigen and driven problems exhibit different responses because of the forced condition on the wavenumber imposed by the exciting plane wave in the driven case.

We expect the OPB-MNT to be a useful tool for the design and characterization of a wide range of LWAs, gratings, reflectors, and metasurfaces.



\newcommand{\BIBdecl}{\interlinepenalty=10000} 



\begingroup
  \bibliographystyle{IEEEtran}
  \bibliography{IEEEabrv,mybib}
\endgroup


\end{document}